\documentclass{aa}

\usepackage{natbib,twoopt}
\usepackage{epstopdf}
\usepackage{epsfig,graphicx,natbib,color}
\usepackage{longtable}
\usepackage[varg]{txfonts}
\usepackage{lscape}
\usepackage{balance}

\newcommand{\felix}[1]{\textsc{F\small{ELIX}}}
\newcommand{\threemodes}[1]{$\nu_0\sim\nu_1+\nu_2$}
\newcommand{\dbv}[1]{the DBV star KIC\,08626021}
\usepackage[breaklinks=true]{hyperref} 
\bibpunct{(}{)}{;}{a}{}{,}             
\makeatletter
  \newcommandtwoopt{\citeads}[3][][]{\href{http://adsabs.harvard.edu/abs/#3}%
    {\def\hyper@linkstart##1##2{}%
     \let\hyper@linkend\@empty\citealp[#1][#2]{#3}}}
  \newcommandtwoopt{\citepads}[3][][]{\href{http://adsabs.harvard.edu/abs/#3}%
    {\def\hyper@linkstart##1##2{}%
     \let\hyper@linkend\@empty\citep[#1][#2]{#3}}}
  \newcommandtwoopt{\citetads}[3][][]{\href{http://adsabs.harvard.edu/abs/#3}%
    {\def\hyper@linkstart##1##2{}%
     \let\hyper@linkend\@empty\citet[#1][#2]{#3}}}
  \newcommandtwoopt{\citeyearads}[3][][]%
    {\href{http://adsabs.harvard.edu/abs/#3}
    {\def\hyper@linkstart##1##2{}%
     \let\hyper@linkend\@empty\citeyear[#1][#2]{#3}}}
\makeatother

\begin{document}

\title{Signatures of nonlinear mode interactions\\ in the pulsating hot 
B subdwarf star KIC\,10139564}
\author{W.~Zong \inst{1,2}
\and S.~Charpinet \inst{1,2}
\and G.~Vauclair \inst{1,2}
}

\institute{Universit\'e de Toulouse, UPS-OMP, IRAP, Toulouse F-31400, France
\and CNRS, IRAP, 14 avenue Edouard Belin, F-31400 Toulouse, France\\ 
\email{[weikai.zong,stephane.charpinet,gerard.vauclair]@irap.omp.eu}}

\date {Received  / Accepted}

\titlerunning{Signatures of nonlinear mode interactions in the pulsating 
hot B subdwarf star KIC\,10139564}
\authorrunning{Zong et al.}

\abstract
{The unprecedented photometric quality and time coverage 
offered by the {\sl Kepler} spacecraft has opened up new opportunities to 
search for signatures of nonlinear effects that affect
oscillation modes in pulsating stars.}
{The data accumulated on the pulsating hot B subdwarf KIC\,10139564 are 
used to explore in detail the stability of its oscillation modes, 
focusing in particular on evidences of nonlinear behaviors.}
{We analyse 38-month of contiguous short-cadence data, concentrating on 
mode multiplets induced by the star rotation and on frequencies forming 
linear combinations that show intriguing behaviors during the course of 
the observations.}
{We find clear signatures that point toward nonlinear effects predicted 
by resonant mode coupling mechanisms. These couplings can induce various 
mode behaviors for the components of multiplets and for frequencies related 
by linear relationships. We find that a triplet 
at 5760\,$\mu$Hz, a quintuplet at 5287\,$\mu$Hz and a ($\ell>2$) 
multiplet at 5412\,$\mu$Hz, all induced by rotation, show clear frequency 
and amplitude modulations which are typical of the so-called intermediate 
regime of a resonance between the components. One triplet at 316\,$\mu$Hz 
and a doublet at 394\,$\mu$Hz show modulated amplitude and constant frequency 
which can be associated with a narrow transitory regime of the resonance. 
Another triplet at 519\,$\mu$Hz appears to be in a frequency lock regime 
where both frequency and amplitude are constant. 
Additionally, three linear combination of frequencies near 6076\,$\mu$Hz 
also show amplitude and frequency modulations, which are likely related to 
a three-mode direct resonance of the type \threemodes{}.}
{The identified frequency and amplitude modulations are the first clear-cut 
signatures of nonlinear resonant couplings occurring in pulsating hot B 
subdwarf stars. However, the observed behaviors suggest that the resonances 
occurring in these stars usually follow more complicated patterns than the 
simple predictions from current nonlinear theoretical frameworks.
These results should therefore motivate further work to develop the theory
of nonlinear stellar pulsations, considering that stars like KIC\,10139564 
now offer remarkable testbeds to do so.}

\keywords{techniques: photometric --
                stars: variables (V361) --
                stars: individual (KIC\,10139564)
               }

\maketitle

\section{Introduction}
Hot B subdwarf (sdB) stars are helium core burning objects that populate 
the so-called Extreme Horizontal Branch (EHB). They are expected to have 
a mass around 0.47 $M_{\odot}$ and are characterized by a very 
thin hydrogen-rich residual envelope containing at most 
$\sim 0.02$ $M_{\odot}$. For this reason, they remain hot 
and compact throughout all their helium core burning evolution, with 
effective temperatures, $T_{\rm eff}$, and surface gravities, $\log g$, 
ranging from 22\,000 K to 40\,000 K and from 5.2 to 6.2, respectively 
\citep{he09,fo12}.

The presence of pulsations in some sdB stars make them good candidates 
for probing their interior with the technique of asteroseismology. 
A first group of nonradial sdB pulsators with periods of a few minutes 
was theoretically predicted by \citet{ch96} and effectively discovered by 
\citet{ki97}. 
These pulsators, now referred to as the V361\,Hya stars,
show low-order, low-degree pressure ($p$-)modes that are driven by 
a $\kappa$-mechanism induced by the partial ionization of iron-group 
elements occurring in the "Z-bump" region and powered-up by radiative 
levitation \citep{ch96,ch97}. Long period oscillations of $\sim1-4$ h 
were later discovered by \citet{gr03}, forming another group of sdB 
pulsators known as the V1093~Her stars. The latter show mid-order gravity 
($g$-)modes driven by the same mechanism \citep{fo03}. 
Hybrid pulsators that show both $p$- and $g$-mode oscillations simultaneously 
have also been reported \citep[e.g., ][]{ss06}. Tight seismic constraints 
have indeed been obtained from the measured frequencies using both types of 
sdB pulsators, in particular based on high-quality photometric data gathered 
from space-borned telescopes \citep[e.g., ][]{ch11b,van10}. However, the 
reason behind the apparent variability of some oscillation modes in sdB stars, 
already noticed from repeated ground based campaigns \citep[e.g., ][]{ki07}, 
has remained poorly understood. 

The temporal variation of oscillation modes in pulsating sdB stars is 
beyond the scope of the standard {\sl linear} nonradial stellar oscillation 
theory in which eigenmodes have a stable frequency and amplitude \citep{un89}. 
These behaviors must be studied within a {\sl nonlinear} framework to interpret 
the modulations. In particular nonlinear resonant mode coupling effects 
are expected to affect some oscillation modes, as noted, e.g., in the helium 
dominated atmosphere white dwarf variable (DBV) star GD~358 \citep{go98}. 
Different types of resonant coupling have been investigated within the 
framework of the amplitude equation (AE) formalism since the 1980's, 
among them the \threemodes{} resonance \citep{dz82,mo85} and the 2:1 
resonance in Cepheid stars \citep{bu86}. The AE formalism was then 
extended to nonadiabatic nonradial pulsations in Eulerian and Lagrangian 
formulations by \citet{go94} and \citet{van94}, respectively. 
A theoretical exploration of specific cases of nonradial 
resonances was developed in \citet{bu95,bu97}, including notably the 
resonance occurring in a mode triplet that is caused by slow stellar rotation
and which satisfies the relationship $\nu_+ + \nu_- \sim 2\nu_0$ , 
where $\nu_0$ is the frequency of the central $m=0$ component.
However, these theoretical developments based on AEs have since considerably
slowed down, in part due to the lack of clear observational data to rely on.

The launch of instruments for ultra high precision photometry from space has 
changed the situation, making it now possible to capture amplitude and/or 
frequency modulations occurring on timescales of months or even years 
that were difficult to identify from ground-based observatories.
It is however from ground based data that \citet{va11} proposed for the 
first time that resonant couplings within triplets could explain the 
long-term variations, both in amplitude and frequency, seen in several 
oscillation modes monitored in the GW\,Virginis pulsator PG\,0122+200, 
through successive campaigns.

The observation of a multitude of pulsating stars, including sdB and white 
dwarf stars, by the {\sl Kepler} spacecraft has open up new opportunities
to indentify and characterize the mechanisms that could modulate the 
oscillation modes. {\sl Kepler} monitored a 105 
deg$^2$ field in the Cygnus-Lyrae region for around four years without 
interruption, thus obtaining unprecedented high quality photometric data 
for asteroseismology \citep{gi10}. These uninterrupted data are particularly 
suited for searching long-term temporal amplitude and frequency 
modulations. 
In the context of white dwarf pulsators, for instance, \citet[hereafter Z16]{zo16} 
found that \dbv{} shows clear signatures of nonlinear effects attributed to 
resonant mode couplings. In this star, three rotational multiplets show 
various types of behaviors that can be related to different regimes 
of the nonlinear resonant mode coupling mechanism. In particular some 
amplitude and frequency modulation timescales are found to be consistant 
with theoretical expectations. This finding suggests that the variations of 
some oscillation modes in sdB stars may also be related to nonlinear 
resonance effects. It is in this context that we decided to search clues of 
similar nonlinear phenomena involving mode interactions in pulsating sdB stars.

Eighteen sdB pulsators have been monitored with {\sl Kepler} 
(see \citealt{os14} and references therein). In this paper, 
we focus on one of them, the star KIC~10139564, which was discovered 
in quarter\,Q2.1 and then continuously observed from Q5.1 to Q17.2. 
A preliminary analysis based on one month of short cadence data originally 
showed that KIC~10139564 is a V361-Hya type (rapid, $p$-mode) sdB pulsator 
featuring also a low-amplitude $g$-mode oscillation \citep{ka10}. 
With extended data, \citet{ba12} detected up to 57 periodicities 
including several multiplets attributed to the rotation of the star. 
These multiplets are characterized by common frequency spacings, both for 
the $p$- and $g$-modes, indicating that KIC~10139564 has a rotation period 
of $25.6\pm1.8$\,d. These authors did not find any radial-velocity 
variations from their dedicated spectroscopy and derived the atmospheric 
parameter values $T_{\rm eff}=31~859$ K and $\log g=5.673$ for this star.
An interesting finding concerning KIC~10139564 is that two of the identified 
multiplets may have degrees $\ell$ greater than 2, a possibility further investigated 
by \citet{bo13}. The detection of several multiplets in this star 
continuously monitored for more than three years makes it a target of choice 
for studying eventual nonlinear resonant mode couplings in sdB stars.

In this study, we show that several multiplets in KIC\,10139564 have indeed 
amplitude and frequency modulations suggesting nonlinear resonant mode 
couplings, which constitutes the first clear-cut case reported for sdB pulsators, 
so far. 
In Sect.\,2, we present the thorough analysis of the frequency 
content of the {\sl Kepler} photometry available on KIC\,10139564, including 
our analysis of the frequency and amplitude modulations identified in several 
multiplets and linear combination frequencies. 
In Sect.\,3, we recall some theoretical background related to nonlinear resonant 
mode couplings, focusing mainly on two types of resonances. 
The interpretation of the observed modulations which may relate to nonlinear 
resonant mode couplings is discussed in Sect.\,4. The summary and conclusion are 
then given in Sect.\,5.

\begin{figure}
\includegraphics[width=8.5cm]{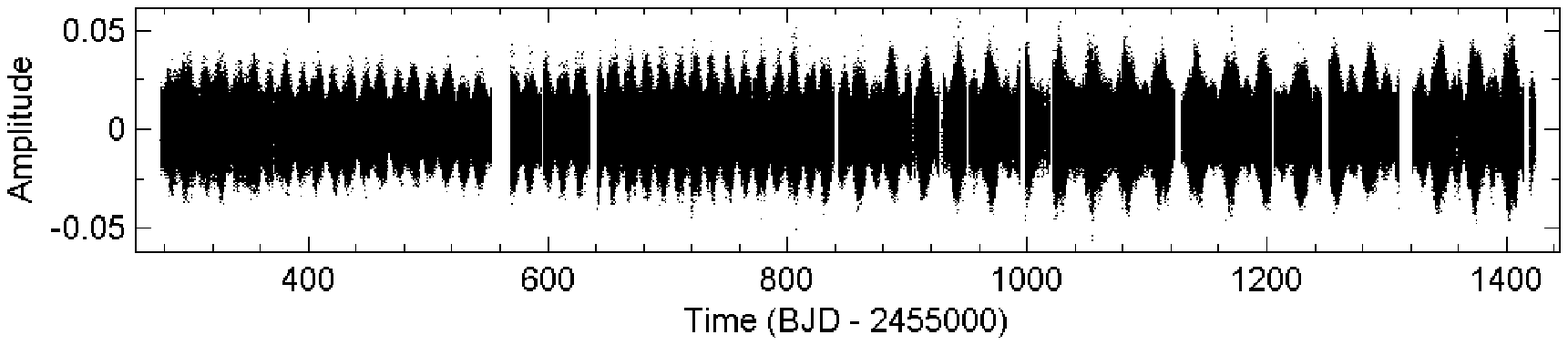}
\includegraphics[width=8.5cm]{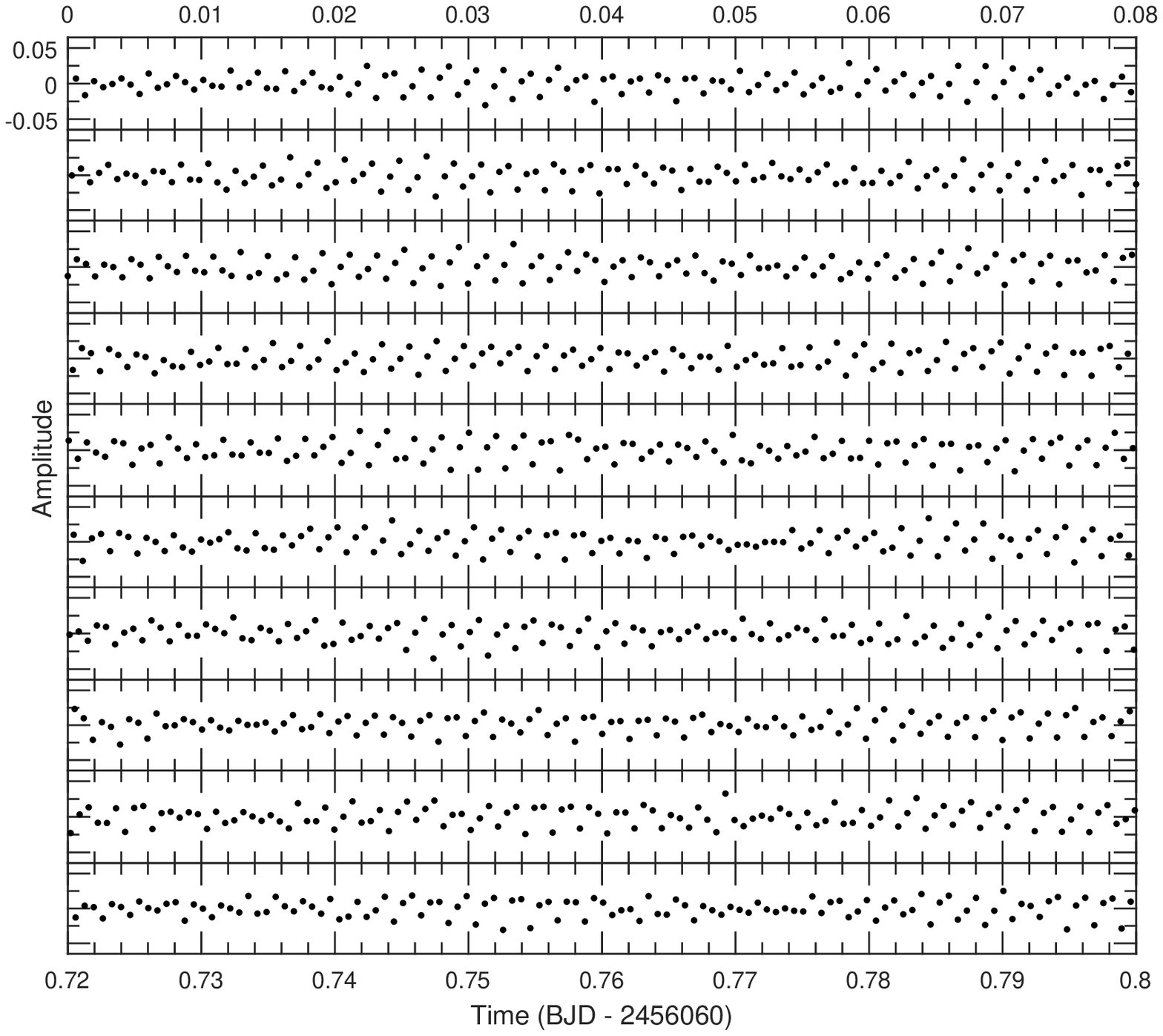}
\caption{{\sl Top panel}: Condensed representation of the full 
{\sl Kepler} light curve of KIC\,10139564 (Amplitude as the 
residual relative to the mean brightness intensity of the star vs time in 
Barycentric Julian Date) covering from Q5.1 to Q17.2 
($\sim 1147.5$ days). {\sl Bottom panel}: 
Close-up view showing 0.8 days of the {\sl Kepler} light curve 
by slices of 0.08 days. At this scale the oscillations 
are clearly apparent. 
\label{lc}}
\end{figure}

\begin{figure*}
\includegraphics[width=17cm]{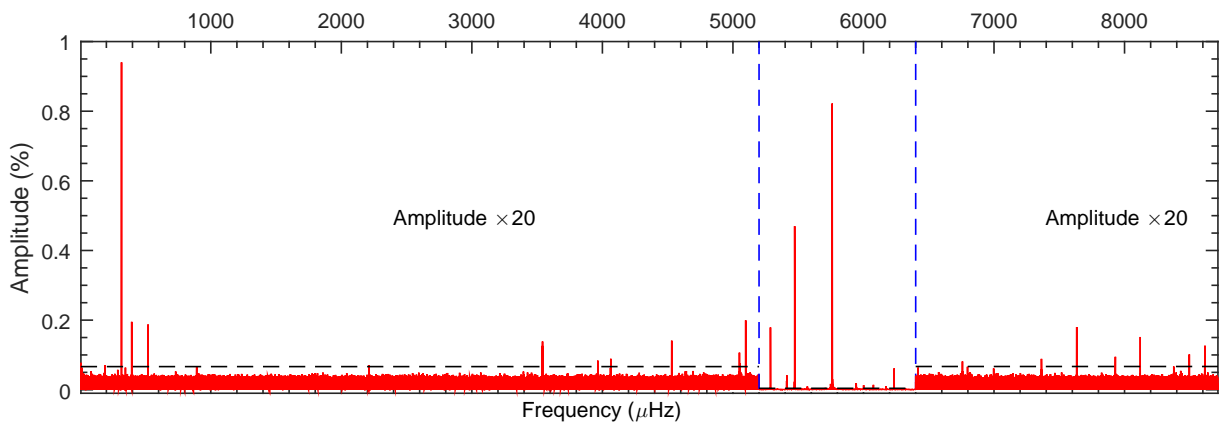}
\caption{Lomb-Scargle Periodogram 
(LSP; Amplitude in \% of the mean brightness vs frequency 
in $\mu$Hz) of the {\sl Kepler} light curve for KIC\,10139564. 
The represented range, up to the Nyquist frequency, covers the 
long-period {\sl g}-mode and the short-period {\sl p}-mode frequency 
domains. The region between the two dashed vertical lines at 5200 and 
6400\,$\mu$Hz is where peaks have the largest amplitudes. However, weaker 
peaks outside of this particular region are present and are 
made visible by scaling up amplitudes by a factor of 20.  
The dashed horizontal line represents the 5.6$\sigma$ detection 
threshold (see text). Some well-known {\sl Kepler} instrumental 
artefacts are present, but can easily be recognized.
\label{lsp}}
\end{figure*}

\section{The frequency content of KIC~10139564 revisited}

\subsection{The $Kepler$ photometry}

The pulsating sdB star KIC\,10139564 was observed by 
{\sl Kepler} in short-cadence (hereafter SC) mode 
during quarter Q2.1 and from Q5.1 to Q17.2 (i.e., until the 
spacecraft finally lost its second inertia reaction wheel and stopped its 
operations). Results based on parts of these data have already been
published in the literature \citep[e.g., ][]{ba12,bo13}.
we obtained the light curves through the $Kepler$ Asteroseismic 
Science Consortium (KASC)\footnote{htpp://astro.phys.au.dk/KASC}. 
These data were processed through the standard {\sl Kepler} 
Science Processing Pipeline \citep{je10}. 
For our purposes, we do not further consider the "short" 
(one month) light curve of Q2.1 which is well disconnected from the main 
campaign and would introduce a large and detrimental gap for our 
upcoming analysis. This leaves us with a nearly contiguous 38-month 
light curve starting from  BJD~2~455~276.5 and ending on BJD~2~456~424 
(which spans $\sim$ 1~147.5 days), with a duty circle of $\sim 89\%$.

The full light curve was constructed from each quarter 
"corrected" light curves, which most notably include 
a correction of the amplitudes taking into account contamination 
by nearby objects (this correction estimates that $\sim 83.2\%$ of the 
light comes from KIC\,10139564). Each quarter light curve 
was individually detrended to correct for residual drifts by performing a 
sixth-order polynomial fit. Then, data points that differ significantly 
from the local standard deviation of the light curve were removed by 
applying a running 3$\sigma$ clipping filter. Note that the latter operation 
decreases slightly the overall noise level in Fourier space, but has 
no incidence on the measured frequencies. 

The fully assembled light curve of KIC\,10139564 is shown in the top panel 
of Fig.\,\ref{lc} while the bottom panel expands a 0.8-day portion 
of the data. Low-amplitude multi-periodic oscillations dominated by 
periodicities of a few minutes are clearly visible. Their presence is 
confirmed in the corresponding Lomb-Scargle Periodgram 
(LSP, Fig.\,\ref{lsp}; \citealt{sc82}). The LSP shows two distinct regions
with significant signal corresponding to $p$-modes at high frequencies and 
$g$-modes at low frequencies. This identifies KIC\,10139564 as a hybrid 
pulsating sdB star \citep{ss06} whose oscillations are however largely 
dominated by $p$-modes. The formal frequency resolution achieved with these
data is $\sim0.010$ $\mu$Hz.

\subsection{Frequency extraction}

A dedicated software, \felix{} (Frequency Extraction for LIghtcurve 
eXploitation) developed by one of us (S.C.), was used to first extract 
the frequency content of KIC\,10139564 down to a chosen detection 
threshold. The latter was established following the same method as in Z16 
(see their Sect.\,2.2), leading also in the present case to a conservative 
5.6$\sigma$ criterion (in practice, we searched down to $\sim 5\sigma$ if 
a frequency is suspected to be part of a multiplet; see below).

The extraction method is a standard prewhithening and nonlinear least 
square fitting technique \citep{de75}, which works with no difficulty in 
the present case. The code \felix{} greatly accelerates and eases 
the application of this procedure, especially for treating very long time-series 
obtained from space with, e.g., CoRoT and {\sl Kepler} \citep{ch10,ch11b}.

\begin{table*} \caption[]{List of frequencies detected in KIC\,10139564
on which we focus our analysis.}
\begin{center}
\begin{tabular}{cccccccccrl}
\hline
\hline
Id.          &Frequency   &$\sigma_f$ & Period  &$\sigma_P$&Amplitude&$\sigma_A$&Phase&$\sigma_\mathrm{Ph}$&S/N &$^\dagger$Comment  \\
&             ($\mu$Hz)   &($\mu$Hz)  &(s)      &(s)       &(\%)     &(\%)                   \\
\hline
&&\\
\multicolumn{3}{l}{Multiplet frequencies:}\\
$f_{39}$ &  315.579243 & 0.000566 & 3168.776214 & 0.005687 & 0.005851 & 0.000596 & 0.2492 & 0.0516 & 9.8   & $T_{2,-1}$\\
$f_{21}$ &  315.820996 & 0.000219 & 3166.350599 & 0.002193 & 0.015155 & 0.000596 & 0.6107 & 0.0199 & 25.4  & $T_{2,0}$\\
$f_{11}$ &  316.066440 & 0.000070 & 3163.891744 & 0.000702 & 0.047276 & 0.000596 & 0.2063 & 0.0064 & 79.3  & $T_{2,+1}$\\
&&\\
$f_{27}$ &  394.027385 & 0.000342 & 2537.894669 & 0.002202 & 0.009667 & 0.000594 & 0.2589 & 0.0312 & 16.3  & $D_{1, 0}$\\
$f_{32}$ &  394.289823 & 0.000397 & 2536.205455 & 0.002555 & 0.008323 & 0.000594 & 0.5123 & 0.0363 & 14.0  & $D_{1, +1}$\\
&&\\
$f_{34}$ &  518.900359 & 0.000437 & 1927.152262 & 0.001624 & 0.007526 & 0.000592 & 0.6648 & 0.0401 & 12.7  & $T_{3, -1}$\\
$f_{28}$ &  519.151796 & 0.000352 & 1926.218898 & 0.001305 & 0.009351 & 0.000592 & 0.9059 & 0.0323 & 15.8  & $T_{3, 0}$\\
$f_{31}$ &  519.402391 & 0.000367 & 1925.289559 & 0.001360 & 0.008964 & 0.000592 & 0.5369 & 0.0337 & 15.2  & $T_{3, +1}$\\
&&\\
$f_{08}$ & 5286.149823 & 0.000053 &  189.173601 & 0.000002 & 0.064784 & 0.000614 & 0.6712 & 0.0047 & 105.4 & $Q_{1,-2}$\\
$f_{10}$ & 5286.561766 & 0.000060 &  189.158861 & 0.000002 & 0.057105 & 0.000614 & 0.4356 & 0.0053 & 92.9  & $Q_{1,-1}$\\
$f_{07}$ & 5286.976232 & 0.000038 &  189.144032 & 0.000001 & 0.088857 & 0.000614 & 0.1202 & 0.0034 & 144.6 & $Q_{1, 0}$\\
$f_{05}$ & 5287.391879 & 0.000019 &  189.129163 & 0.000001 & 0.179339 & 0.000615 & 0.3374 & 0.0017 & 291.8 & $Q_{1,+1}$\\
$f_{06}$ & 5287.805883 & 0.000029 &  189.114355 & 0.000001 & 0.119329 & 0.000615 & 0.7941 & 0.0025 & 194.2 & $Q_{1,+2}$\\
&&\\
$f_{22}$ & 5410.701146 & 0.000234 &  184.818931 & 0.000008 & 0.014871 & 0.000627 & 0.9524 & 0.0203 & 23.7  & $M_{1, 0}$\\
$f_{67}$ & 5411.143448 & 0.000958 &  184.803824 & 0.000033 & 0.003637 & 0.000627 & 0.4591 & 0.0830 & 5.8   & $M_{1, 0}$\\
$f_{13}$ & 5411.597301 & 0.000136 &  184.788325 & 0.000005 & 0.025636 & 0.000627 & 0.6770 & 0.0118 & 40.9  & $M_{1, 0}$\\
$f_{15}$ & 5412.516444 & 0.000185 &  184.756944 & 0.000006 & 0.018812 & 0.000627 & 0.8925 & 0.0160 & 30.0  & $M_{1, 0}$\\
$f_{12}$ & 5413.389096 & 0.000084 &  184.727161 & 0.000003 & 0.041339 & 0.000627 & 0.4037 & 0.0073 & 65.9  & $M_{1, 0}$\\
$f_{19}$ & 5413.814342 & 0.000222 &  184.712651 & 0.000008 & 0.015718 & 0.000627 & 0.7225 & 0.0192 & 25.1  & $M_{1, 0}$\\
&&\\
$f_{01}$ & 5760.167840 & 0.000005 &  173.606052 & $\dots$  & 0.825132 & 0.000761 & 0.0744 & 0.0004 &1084.9 & $T_{1,-1}$\\
$f_{03}$ & 5760.586965 & 0.000008 &  173.593421 & $\dots$  & 0.554646 & 0.000761 & 0.6388 & 0.0005 & 729.3 & $T_{1,0}$\\
$f_{02}$ & 5761.008652 & 0.000007 &  173.580715 & $\dots$  & 0.567034 & 0.000761 & 0.5845 & 0.0005 & 745.5 & $T_{1,+1}$\\
&&\\
\multicolumn{3}{l}{Linear combination frequencies $C_1$:}\\
$f_{23}$ & 6076.234996 & 0.000252 &  164.575597 & 0.000007 & 0.014360 & 0.000650 & 0.7906 & 0.0210 & 22.1  &  $f_{11}+f_{01}$ \\
$f_{35}$ & 6076.408232 & 0.000510 &  164.570905 & 0.000014 & 0.007091 & 0.000650 & 0.7821 & 0.0426 & 10.9  &  $f_{21}+f_{03}$ \\
$f_{74}$ & 6076.650684 & 0.001120 &  164.564338 & 0.000030 & 0.003225 & 0.000650 & 0.5520 & 0.0937 & 5.0   &  $f_{11}+f_{03}$ \\

\hline
\end{tabular}
\end{center}
\label{t1}
\tablefoot{
\tablefoottext{$\dagger$}{The first subscript is the identity of the multiplet and the second one indicates the value of $m$. The $m$-values for the $\ell>2$ 
multiplet $M_1$ are not provided, as the degree $\ell$ is not known.}
}
\end{table*}

We provide in Table~\ref{t2} (see Appendix) a list of all the extracted 
frequencies with their fitted attributes (frequency in $\mu$Hz, period in second, 
amplitude in percent of the mean brightness, phase relative to a reference 
time $t_0$, and signal-to-noise ratio) and their 
respective error estimates ($\sigma_f$, $\sigma_P$, $\sigma_A$, and 
$\sigma_\mathrm{Ph}$). For convenience, because in this study we focus on
a particular subset of the observed frequencies, we repeat some of these 
information in Table~\ref{t1} for the relevant modes. The "Id." column in both 
tables uniquely identify a detected frequency with the number indicating the 
rank by order of decreasing amplitude.

We have detected 60 clear independent frequencies that comes out well 
above the 5.6$\sigma$ detection threshold (Table~\ref{t2}), of 
which 29 frequencies consist of three triplets, one doublet, one 
quintuplet and two incomplete multiplets with $\ell>2$ (Table~\ref{t1}). 
We also detect another three frequencies that appear as significant but are 
linked to other frequencies through linear combinations. 
Five additional "forests" of frequencies, each containing many close 
peaks in a very narrow frequency range, are detected in the 5400--6400\,$\mu$Hz
region. We also prewhitened 14 frequencies whose amplitudes are above 
5.0$\sigma$ but below 5.6$\sigma$ which, we suspect, are real pulsations. 
Our well-secured extracted frequencies agree well with the
independent analysis of \citet{bo13}, but we detect a few more low-amplitude 
frequencies because the data that we consider here cover about one more year. 
We do not investigate further these "forests" of frequencies 
(G1--G5, see Table~\ref{t2} in Appendix) that show very complicated 
structures. These were 
discussed in \citet{ba12}. We point out that our extracted frequencies 
may differ in amplitude compared with the work of \citet{bo13} because 
some of these frequencies have variable amplitudes. 

\begin{figure*}[htbp]
\centering
\includegraphics[width=8.5cm]{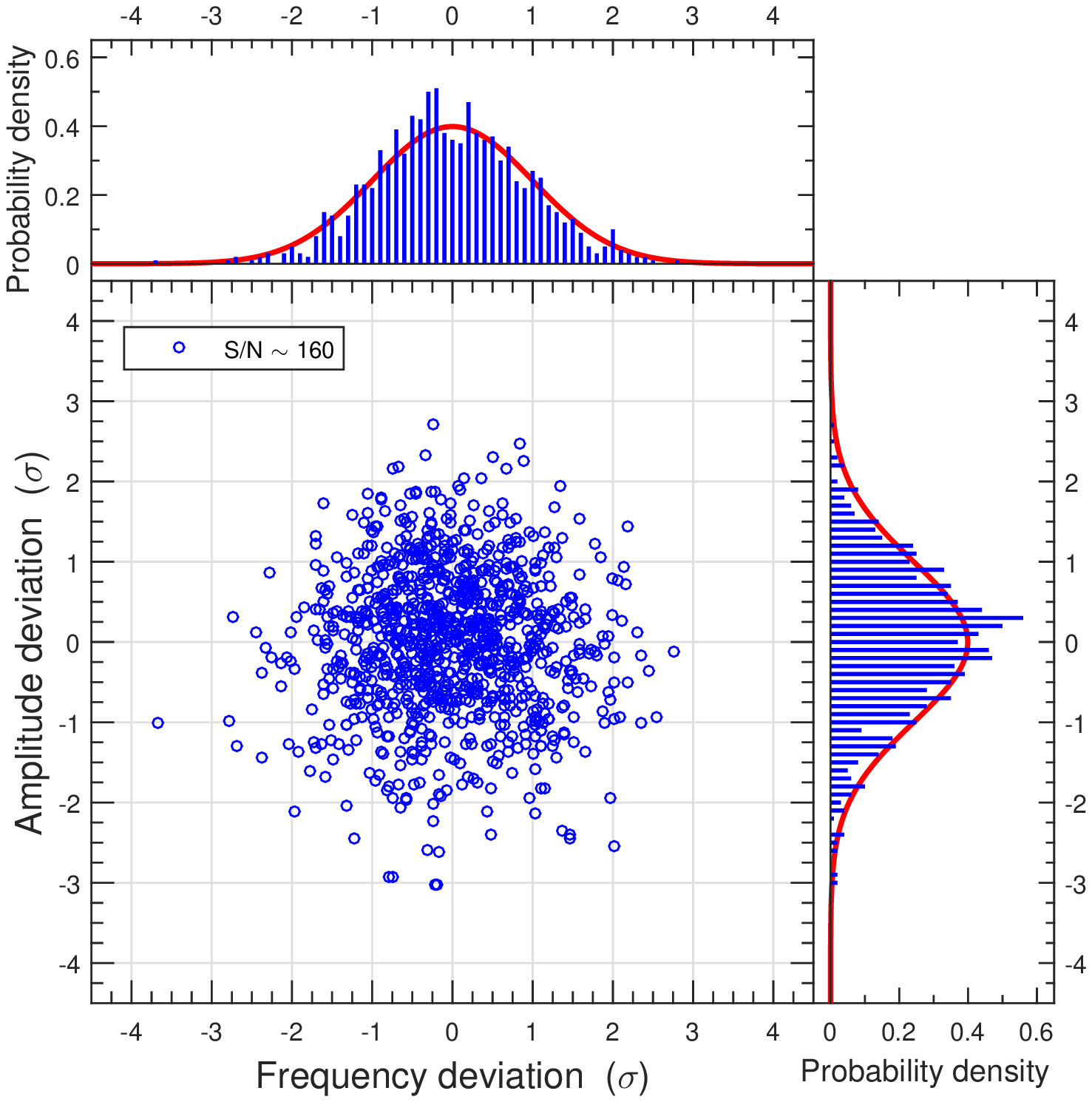}
\includegraphics[width=8.5cm]{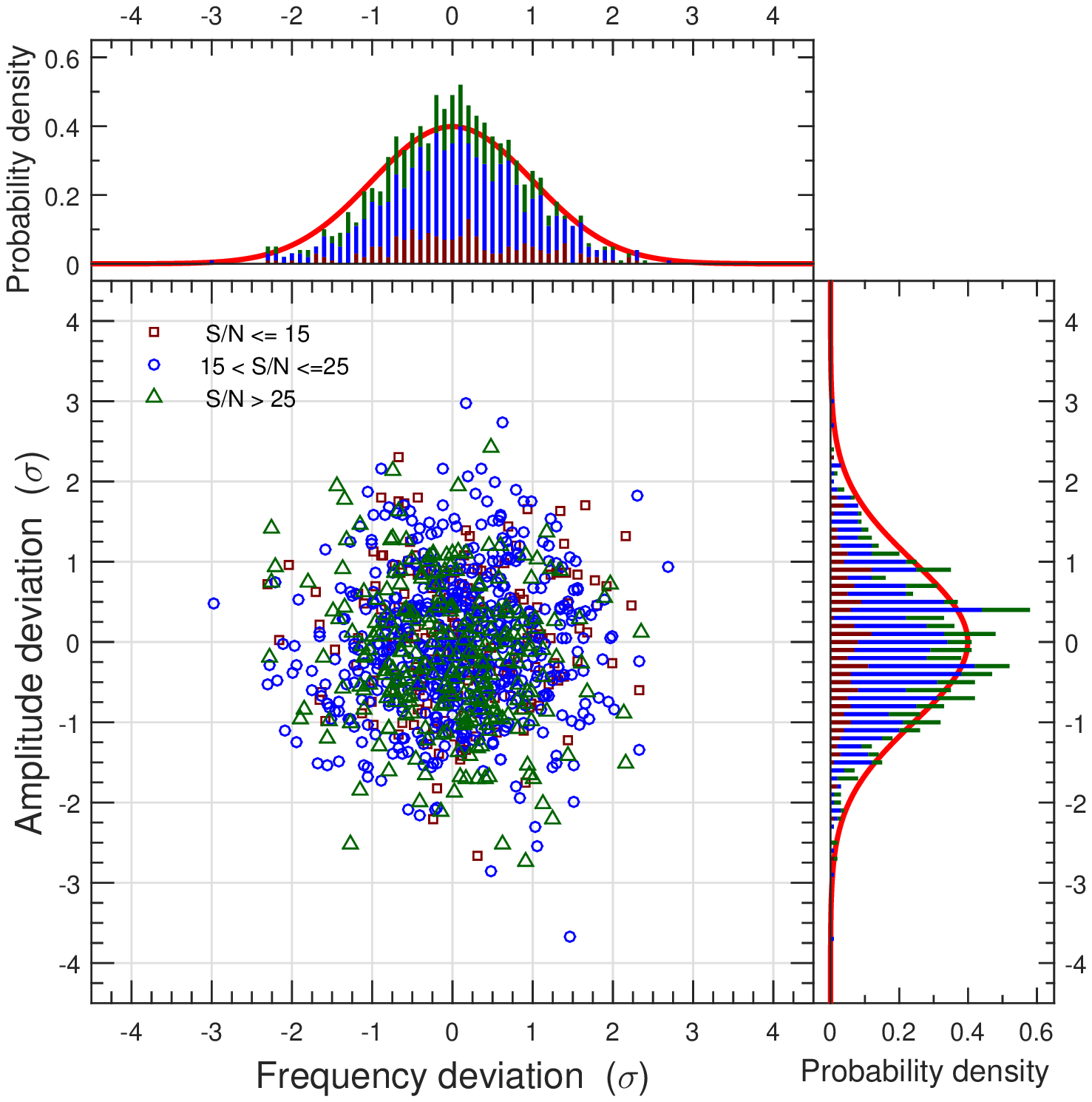}
\caption{{\sl Left panel}: 2-D~distribution of the frequency and amplitude 
deviations between the prewhitened and the injected values for 1\,000 
artificial modes of constant amplitude. S/N denotes the signal-to-noise 
ratio of the injected signals and the deviations have been normalized by 
the 1$\sigma$ error, $\sigma_A$ and $\sigma_f$, derived from the 
prewhitenning procedure implemented in the code \felix{}. 
The 2-D distribution is also projected into 1-D histograms 
(frequency and amplitude) to be compared with the Normal Distribution, 
$\mathcal{N}(0,1)$ plotted as a red solid curve. 
{\sl Right panel}: Same as above but for 1\,000 modes with random amplitudes. 
The injected modes are divided into three groups of S/N ratios in the 
ranges $[5,15]$, $(15, 25]$, and $(25, 60]$, respectively (represented by 
three different colors and symbols).
\label{afd}} 
\end{figure*}

\subsection{Error estimates on frequencies and amplitudes}

Before proceeding further in our analysis, we briefly discuss our 
quantitative evaluation of the uncertainties associated with the measured
frequencies and amplitudes given in Table~\ref{t1} and Table~\ref{t2}. 
The reliability of these error estimates is particularly important when it 
comes to discuss amplitude and frequency variations with time, in particular 
to assess if these are significant or not. 

With \felix{}, errors are estimated following the formalism proposed by
\citet{mont99}, with the particularity, however, that $\sigma_A$, the error 
on the amplitude of a mode, is measured directly in the Lomb-Scargle 
periodogram. A window around each frequency is chosen and the median 
value of the amplitudes in that frequency range defines $\sigma_A$. The 
relations given in \citet{mont99} are then used to compute the other errors,
in particular $\sigma_f$, the error on the measured frequency. In order to 
test that this procedure is correct and does not largely under or 
overestimate the true errors, we conduct two Monte-Carlo experiments.

We first construct an artificial light curve covering about 200 days (similar 
to the time baseline of each light curve pieces considered in the next subsection) 
with the same SC-mode sampling provided by {\sl Kepler} in which we add white 
random gaussian noise. We further inject in this light curve 1\,000 sinusoidal signals 
with the same amplitude (S/N~$\sim160$) but of frequency increasing by steps of 
$\sim8.2\,\mu$Hz per signal. {In practice, a random frequency shift of a few tenth 
$\mu$Hz is performed on each injected frequency in order to reduce the number 
of harmonics and linear combinations.} 
The generated time series is then analysed with our code \felix{} that 
extracts and measures each signal and evaluates the uncertainties associated
to the measured frequencies and amplitudes ($\sigma_A$ and $\sigma_f$). 
Since the true values of these quantities are perfectly known from the 
signals we injected, the real distribution of the deviations between measured 
(prewhitenned) values and true values can be evaluated. For that purpose, 
we define the frequency and amplitude deviations normalized by their 
1$\sigma$ errors (as estimated with the code \felix{} from the procedure 
described above), $\Delta{}f = (f_{\rm pre} - f_{\rm inj})/\sigma_f$ and 
$\Delta{}A = (A_{\rm pre} - A_{\rm inj})/\sigma_A$, where the subscripts indicate 
the prewhitened value and the injected one, respectively. A variant of this
test is also performed by again injecting 1\,000 sinusoidal signals, but this
time with random amplitudes (instead of constant ones) chosen in the 
S/N $\in (5, 60)$ range. This second test allows us to check also the 
reliability of our error estimates as a function of amplitude, 
considering that $\sigma_f$ in particular depends on the mode S/N ratio 
($\sigma_f$ increases when S/N decreases).

Figure\,\ref{afd} shows the results obtained in both cases. The 2-D 
distributions of the frequency and amplitude deviations are well confined 
within 3$\sigma$. Moreover, the associated 1-D histograms show that 
for both quantities, the measured deviations closely follow the Normal 
Distribution, $\mathcal{N}(0,1)$, plotted as a red solid curve. Only a few 
data points fall outside the $[-3\sigma, +3\sigma]$ range (within which 
99.73\% of the measurements should be for the normal distribution, 
$\mathcal{N}(0,1)$). This is the behavior we expect for an accurate 
determination of the error estimates, $\sigma_A$ and $\sigma_f$, with 
the code \felix{}. Hence, these tests demonstrate that error values 
derived in our frequency analysis are robust.  

\subsection{Amplitude and frequency modulations}

\begin{figure*}
\centering
\includegraphics[width=15.5cm]{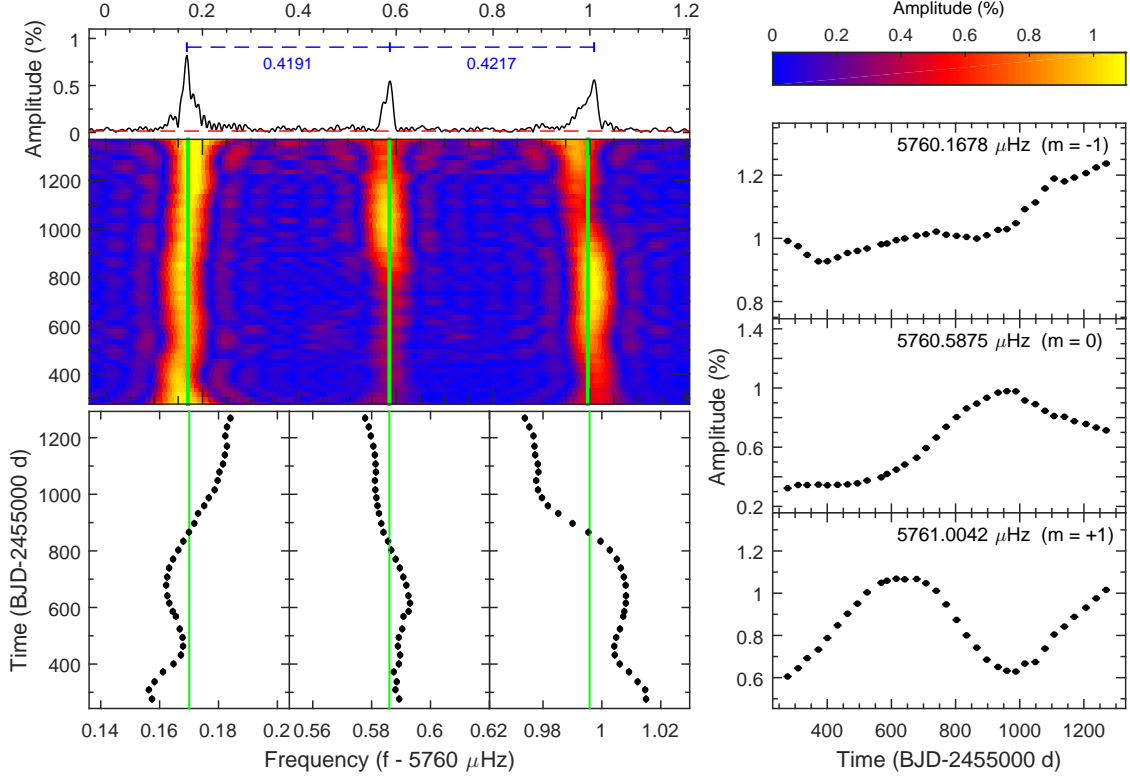}
\caption{Frequency and amplitude modulations in the $T_1$ {\sl p}-mode triplet near 
5760\,$\mu$Hz. {\sl Top-left panel} presents the fine structures of the well defined 
triplet with near symmetric frequency spacings. The dashed horizontal line in red 
represents the 5.6\,$\sigma$ detection threshold. 
{\sl Middle-left panel} shows the sliding Lomb-Scargle Periodogram 
(sLSP giving the amplitude in \% as a function of frequency in $\mu$Hz and time 
in days) of the triplet as a whole.{\sl Bottom-left panel} shows 
expanded views around the average frequency (the solid vertical 
lines, also in the middle left panel) of each component, 
obtained from prewhitening subsets 
of the data, thus measuring precisely the frequencies, as a function 
of time. {\sl Right panel} provides the measured amplitudes 
as a function of time obtained for each subset of data 
(see text for details). Note that the errors for each measurement is 
smaller than the symbol itself.\label{p1}}
\end{figure*}

\begin{figure*}
\begin{center}
\includegraphics[width=15.5cm]{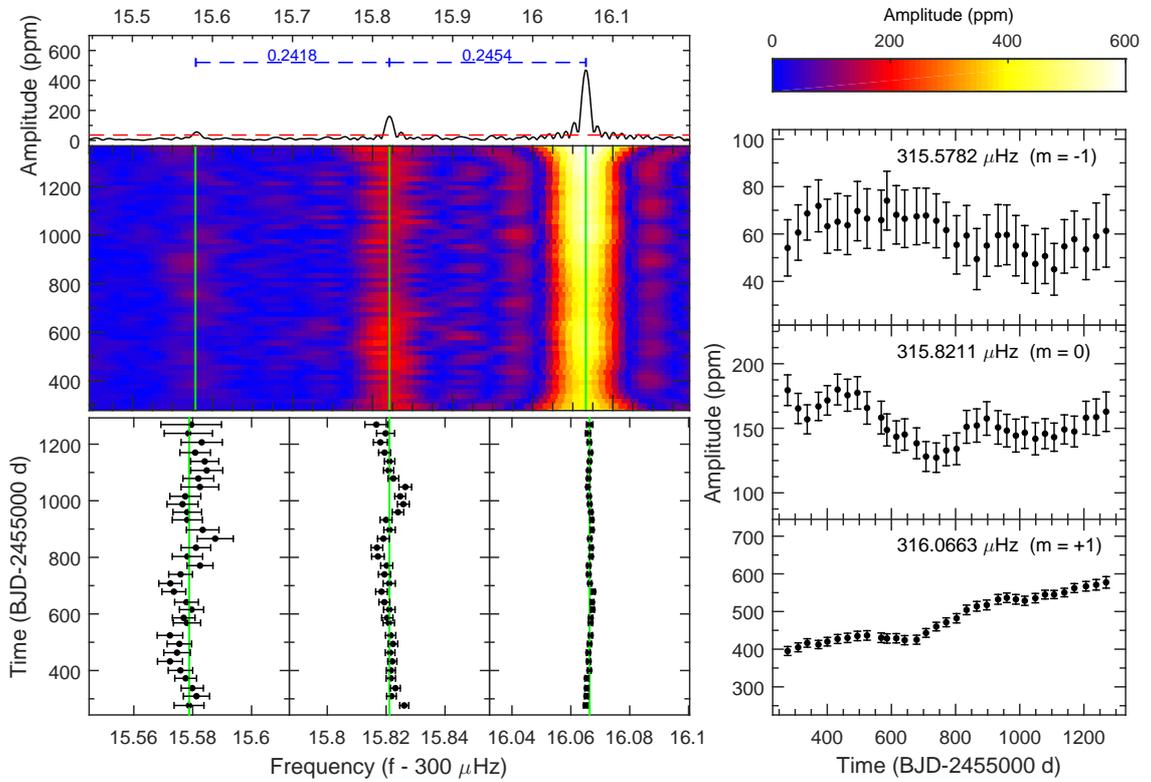}
\end{center}
\caption{Same as Fig.\,\ref{p1} but for the $T_2$ {\sl g}-mode 
triplet near 316\,$\mu$Hz.
\label{g1}}
\end{figure*}

\begin{figure*}
\begin{center}
\includegraphics[width=15.5cm]{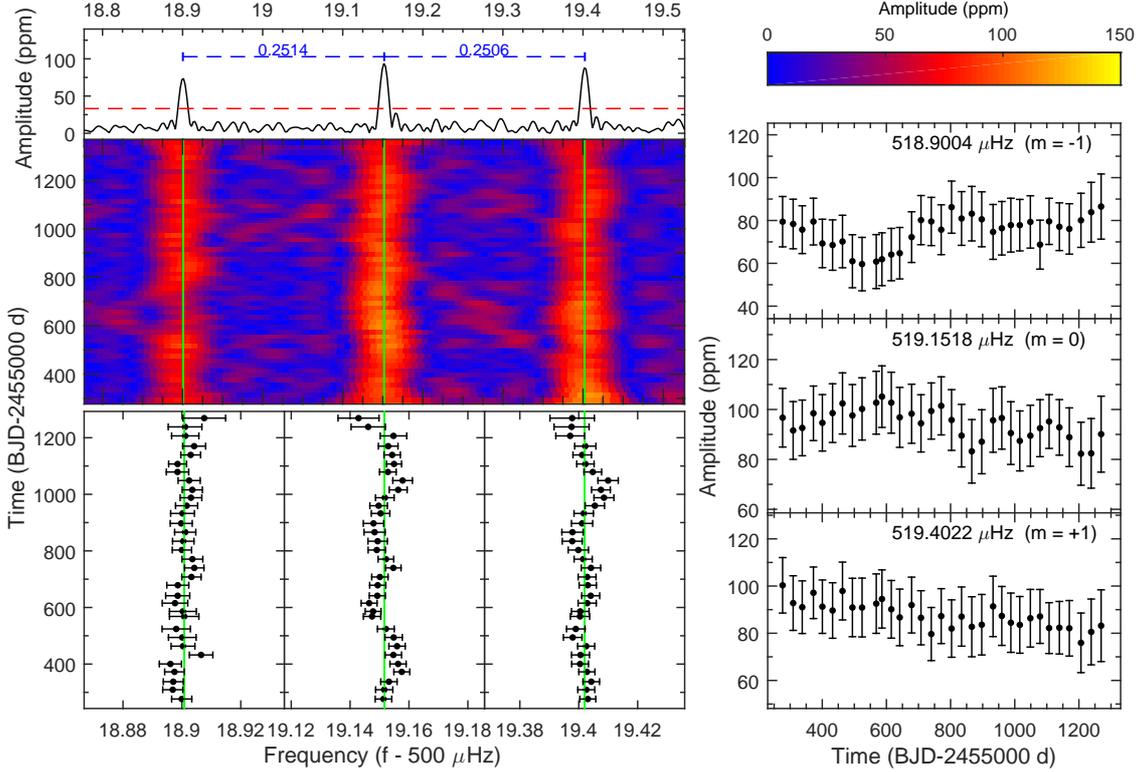}
\end{center}
\caption{Same as Fig.\,\ref{p1} but for the $T_3$ {\sl g}-mode 
triplet near 519\,$\mu$Hz. 
\label{g3}}
\end{figure*}

\begin{figure*}
\sidecaption
\includegraphics[width=12cm]{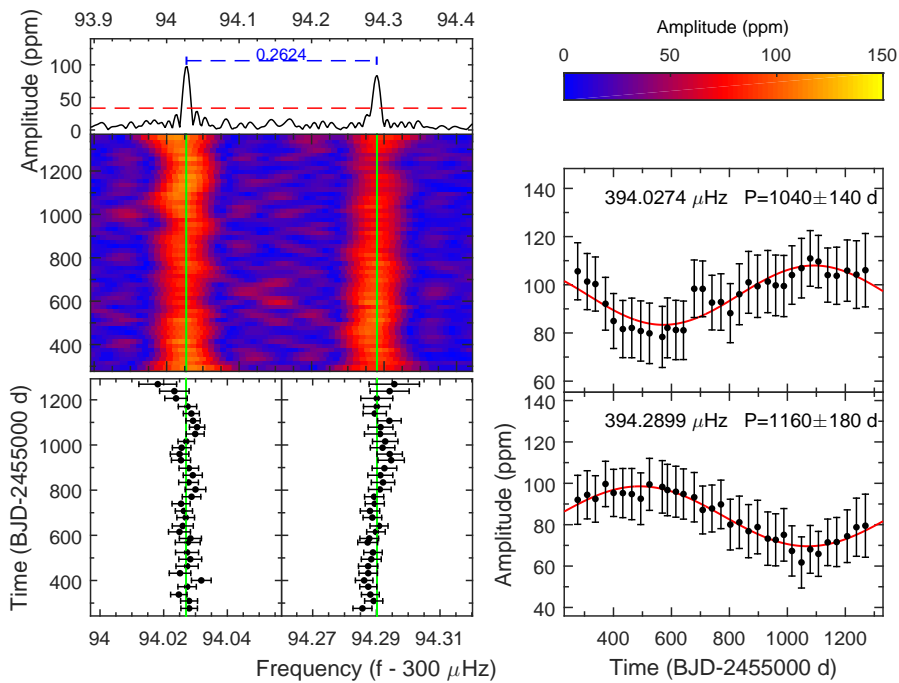}
\caption{Same as Fig.\,\ref{p1} but for the $D_1$ {\sl g}-mode 
doublet near 394\,$\mu$Hz. The solid curves in right panel show 
the best fit of a pure {\sl Sine} wave to the amplitude modulations.
\label{g2}}
\end{figure*}

\begin{figure*}
\begin{center}
\includegraphics[width=17.5cm]{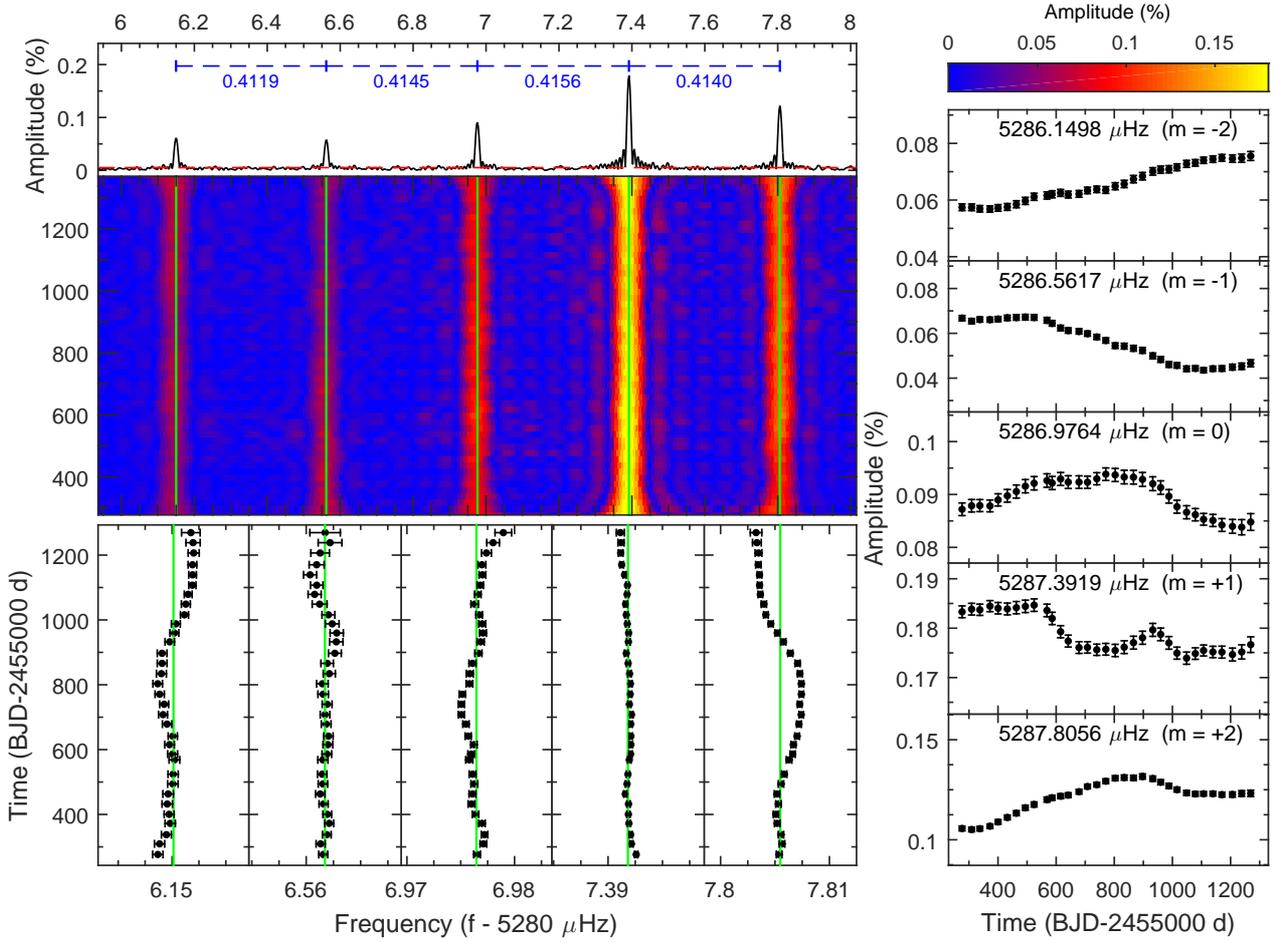}
\end{center}
\caption{Same as Fig.\,\ref{p1} 
but for the $Q_1$ {\sl p}-mode quintuplet near 5287\,$\mu$Hz.
\label{q1}}
\end{figure*}

\begin{figure*}
\begin{center}
\includegraphics[width=17.5cm]{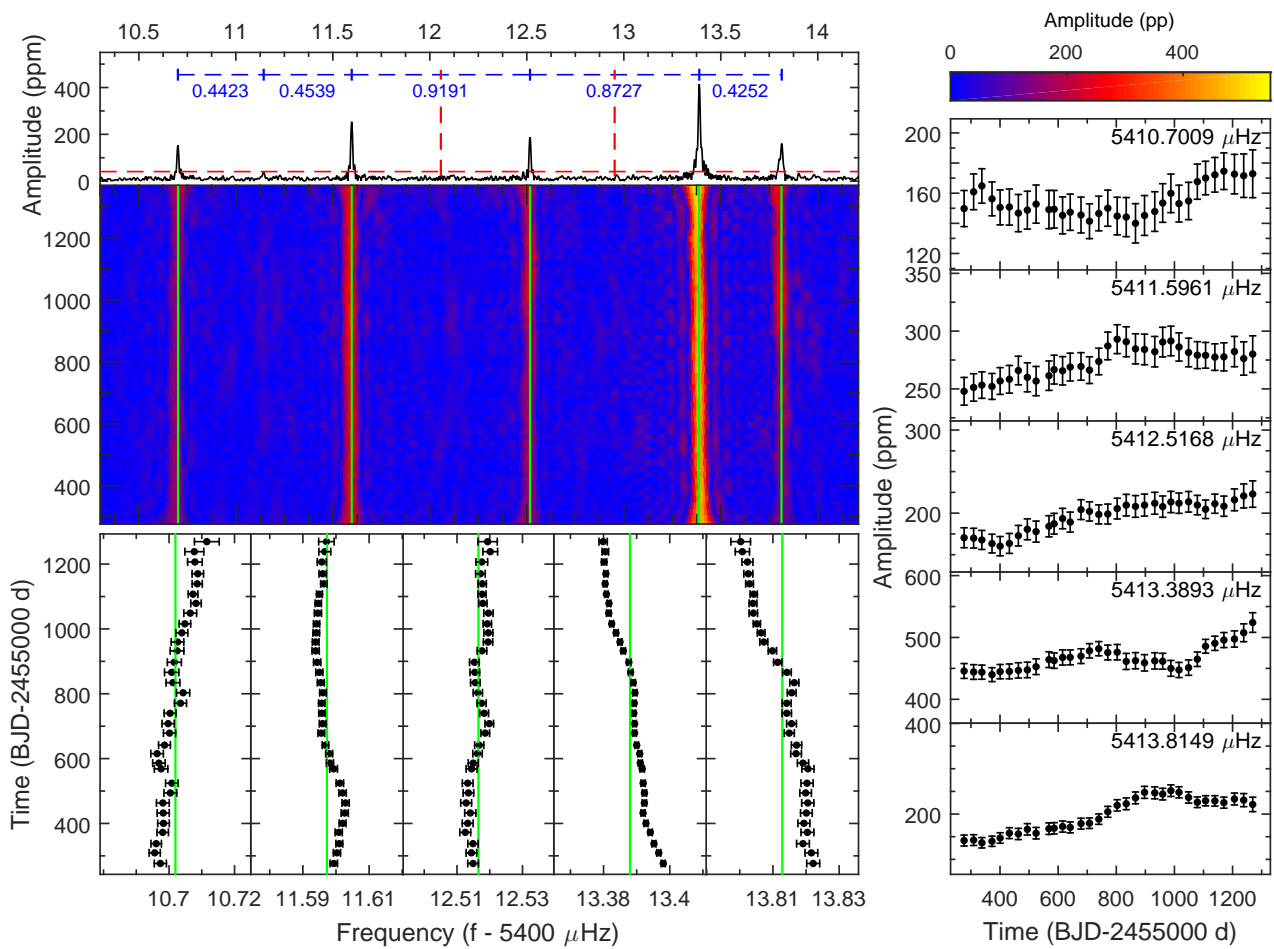}
\end{center}
\caption{Same as Fig.\,\ref{p1} 
but for the $\ell>2$ {\sl p}-mode multiplet $M_1$ near 5413\,$\mu$Hz. 
Note that at least three components are missing in this multiplet 
and the red vertical dashed lines indicate the expected position 
for two of them.
\label{s1}}
\end{figure*}

From now on, we concentrate our discussion on the six multiplets, which
include three triplets $T_1$, $T_2$ and $T_3$, one doublet $D_1$, one 
quintuplet $Q_1$ and a likely $\ell=4$ multiplet $M_1$ (see again 
Table~\ref{t2}). Interestingly, three of these multiplets 
($T_2$, $D_1$, and $T_3$) involve $g$-modes, while the others 
($Q_1$, $M_1$, and $T_1$) are $p$-modes. We also examine three linear 
combination frequencies ($C_1$). The fine structures of the multiplets 
are shown in the left top 
panels of Figs.\,\ref{p1}, \ref{g1}, \ref{g3} and \ref{g2}-\ref{s1}. 
The average frequency spacing between the components of these well-defined, 
nearly symmetric multiplets is $\sim 0.25$\,$\mu$Hz for the $g$-modes and 
0.423\,$\mu$Hz for the $p$-modes, thus suggesting that the $g$-modes are 
dipoles ($\ell=1$) in a star rotating rigidly with a period of $\sim 26$ days.

\begin{figure}
\includegraphics[width=8cm]{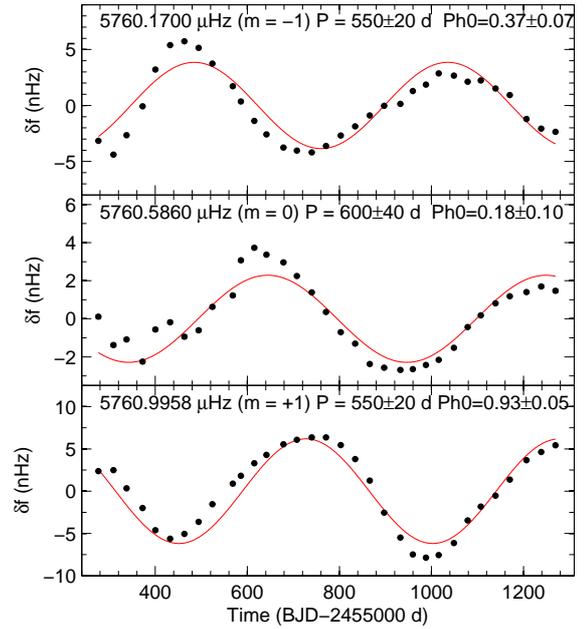}
\caption{Frequency modulations after removing the long-term 
trend in the $T_1$ triplet by applying a second-order polynomial 
fit. The 
solid curves represent the best fits of one pure {\sl Sine} 
wave to the frequency modulations. The associated formal 
errors for the periods and phases are also estimated.
\label{pr1}}
\end{figure}

\begin{figure*}
\centering
\includegraphics[width=14cm]{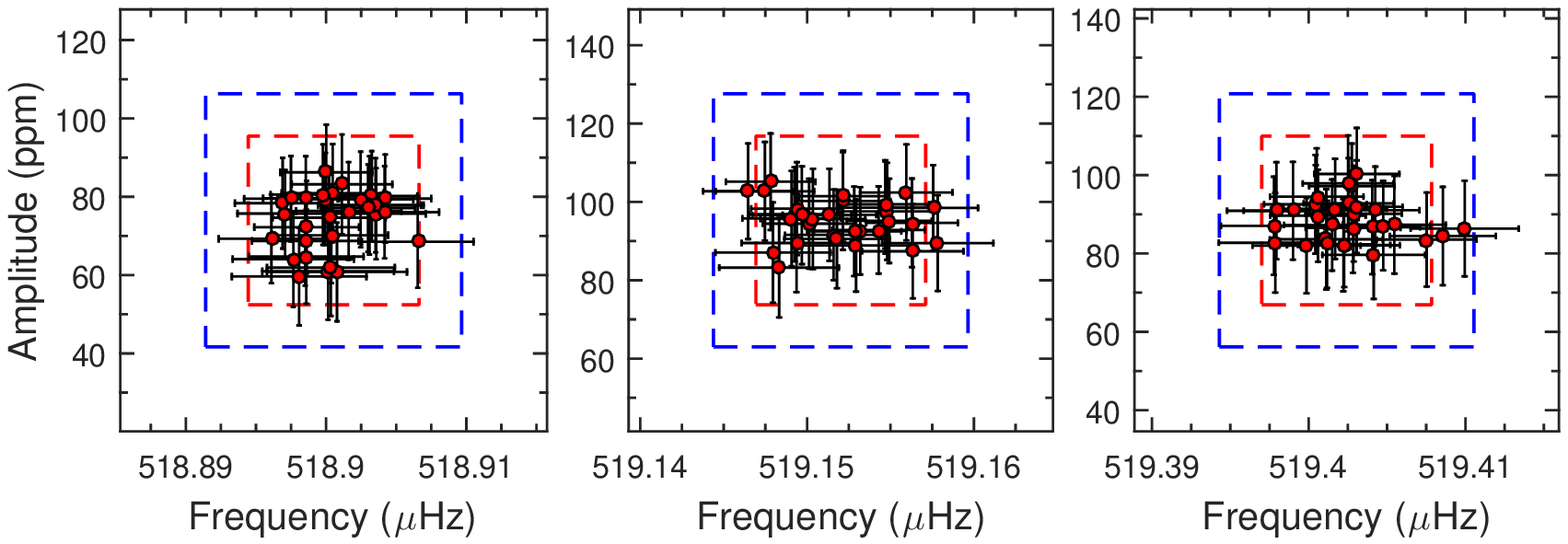}
\caption{Frequency and amplitude scattering for the three components 
forming the {\sl g}-mode triplet $T_3$ around their averaged values. 
The red and blue dashed rectangles indicate the 2$\sigma$ 
and 3$\sigma$ error boxes, respectively. All the data points are within 
3$\sigma$.
\label{sg3}}

\end{figure*}

In order to investigate the time variability of these oscillation modes 
and their relationships, we used our software \felix{} to 
compute sliding Lomb-Scargle Periodograms (sLSP) of the 
data set. This method constructs time-frequency diagrams by 
filtering in only parts of the data set as a function of time. 
We chose a filter window of 200-day width moved along the entire light 
curve by time steps of 20 days. This ensures a good compromise, for our 
purposes, between the frequency resolution (to resolve close 
structures of peaks in each LSP), time resolution, and signal-to-noise. 
The sLSP offers an overall view of the amplitude and frequency variations 
that may occur for a given mode (see, e.g., the middle left panel of 
Fig.\,\ref{p1}). As a complementary (and more precise) technique, we also 
extracted the frequencies (through prewhitening and nonlinear least square 
fitting) in various parts of the light curve. The 
38-month light curve of KIC\,10139564 was divided into 32 time
intervals, each containing 9 months of {\sl Kepler} data (for 
the purposes of precision in the measurements) except for the 
last 3 intervals at the end of the observations. This second 
approach provides a measure of the (averaged) frequencies and 
amplitudes at a given time, along with the associated errors (see, 
e.g., the right and bottom left panels of Fig.\,\ref{p1}).

Figure\,\ref{p1} shows the amplitude and frequency modulations 
for the three components forming the triplet $T_1$ near 5760\,$\mu$Hz. 
As mentioned already, the top-left panel shows the triplet as revealed
by the full data set with components nearly equally spaced in frequency.
We note, however, that this spacing is not strictly symmetric, with a 
difference (or "frequency mismatch") of 0.0026\,$\mu$Hz.
Frequency variations with time are illustrated by the sLSP diagram 
in the middle-left panel where the color scale represents the amplitude of 
the modes. An expanded view centered on the average frequency of each 
component is then provided in the bottom-left panel while the amplitude 
behavior with time for each component is shown in the right panel. The 
latter two are obtained from prewhitening parts of the light curve as 
described above.

From the sLSP diagram, we find that both the 
amplitudes and frequencies have varied during the {\sl Kepler} 
observations. These variations are more clearly seen in the 
bottom-left and right panels. The side components both show 
suggestions of a quasi-periodic modulation in frequency and 
evolve in antiphase. We also note a long timescale trend as the frequencies 
of the two side components gradually approach toward each other, as well as 
toward the central component. In order to filter out these trends, we
applied a parabolic fit to each component, leaving the remaining 
signature of the quasi-periodic modulation of the frequencies 
(see Fig.\,\ref{pr1}). In the process, we find that the two side components
had frequencies about 0.06\,$\mu$Hz closer to each other at the end of the 
run compared to the beginning of the observations. Figure\,\ref{pr1} shows
that the data almost cover two cycles of the quasi-periodic frequency variations.
While clearly not strictly sinusoidal, although not very far from it, if we 
fit the closest pure sine wave to each curve, we find that all have a very similar 
(quasi-)period of $\sim 570$ days.
The variations for the side components (retrograde and prograde modes) 
are clearly in antiphase. 
For the amplitude variations, we also find suggestions of a quasi-periodic 
modulation for the central and prograde components. The retrograde mode 
for its part has a more regular amplitude evolution (increase) during the 
course of the observations.

Figure\,\ref{g1} illustrates the amplitude and frequency modulations 
for the {\sl g}-mode triplet $T_2$ near 316\,$\mu$Hz using the same 
presentation as in Fig.\,\ref{p1}. In this case, the triplet shows a 
very small (but significant) asymmetry of 0.0036\,$\mu$Hz.
The frequencies appear to be stable over the 38-month {\sl Kepler} 
observations. The amplitude is essentially constant for the retrograde 
($m=-1$) mode while the other two components display some variations. 
The central
one may show a small oscillatory behavior, but more precise measurements 
would be needed to really confirm that trend. The prograde ($m=1$) mode has 
its amplitude rising continuously throughout the observations, from 400\,ppm 
up to about 600\,ppm.

The amplitude and frequency variations of the {\sl g}-mode 
triplet $T_3$ near 519\,$\mu$Hz are shown in Fig.\,\ref{g3}. 
In this triplet, which is almost perfectly symmetric, the three 
components have stable frequencies and amplitudes within the quoted 
uncertainties. This stability is further illustrated with Figure\,\ref{sg3} 
that shows the scattering of the measured frequencies and amplitudes for these
modes from all the data chunks considered throughout the entire light curve. 
Almost all measurements are indeed confined within 2$\sigma$ around their 
average values (and all are within 3$\sigma$). It is interesting to note that 
the triplet $T_3$ therefore has different characteristics compared to the 
two triplets $T_1$ and $T_2$.

Figure\,\ref{g2} shows the amplitude and frequency modulations 
for the {\sl g}-mode doublet $D_1$ near 394\,$\mu$Hz. The frequencies 
of each component forming the doublet appear to be stable over the 
38-month {\sl Kepler} observations, while the amplitudes show very suggestive
indications of quasi-periodic variations. We find that the amplitude 
modulations of the two components have very similar periods, about 1100 
days, as illustrated by the best-fit sine waves to the data (The red 
solid curves in right panel of Fig.\,\ref{g2}). Hence the available {\sl Kepler}
data just cover about one cycle of this variation, but it is remarkable that 
almost all the amplitude measurements match very closely the fitted sine 
curves. This estimated period is almost twice the period of modulations 
occurring in the main triplet $T_1$.
Moreover, we clearly see that the amplitudes of the two 
components evolve almost in antiphase during the observing run.

Figure\,\ref{q1} shows the amplitude and frequency modulations 
for the {\sl p}-mode quintuplet $Q_1$ near 5287\,$\mu$Hz. In 
this complete quintuplet, the $m=\pm 2$ modes and possibly the $m=0$ central
component show significant frequency variations. The other modes, 
with $m=\pm 1$, have frequencies which are rather stable (with only marginal 
fluctuations) over the entire observation run. In constrast, the amplitudes 
for all the modes in the quintuplet vary with patterns that cannot clearly be 
connected to periodic modulations, based on the available data. Of course, 
quasi periodic  modulations with a timescale longer than twice the present 
{\sl Kepler} observation cannot be ruled out.
We also note that the frequency variations of the $m=-2$ and $m=+2$ components 
and the amplitude modulations of the $m=-2$ and $m=-1$ components are roughly 
in antiphase during the observation.

Figure\,\ref{s1} shows the amplitude and frequency modulations 
for the $\ell>2$ {\sl p}-mode multiplet $M_1$ near 5413\,$\mu$Hz. 
The $\ell$-value for this group of modes is not clearly assessed yet, 
but a plausible interpretation is that it corresponds to  
an $\ell=4$ nonuplet \citep{ba12} with three undetected components and 
one component barely visible in the LSP of the full data set (see top-left 
panel of Fig.\,\ref{s1}) but which is too low in amplitude to be studied in 
subsets of the light curve.
Some of the frequencies and amplitudes of the five clearly visible modes in 
this multiplet show significant variations during the 38 months of 
{\sl Kepler} observation. In particular, the frequencies of the side 
components drifted toward each other by $\sim 0.032$\,$\mu$Hz from the 
beginning to the end of the run. This trend may be comparable to that 
observed in the side components of the $T_1$ triplet (see Fig.\,\ref{p1}). 
Moreover, the same phenomenon also occurs for some modes observed in the 
long-period-dominated sdB pulsator KIC\,2697388 (M.D. Reed, private 
communication).

In addition to the six multiplets discussed above, we identified other 
possible multiplets in the data, such as six modes near 5571\,$\mu$Hz (see
Table\,\ref{t2} in Appendix), 
in which however most of the components have amplitudes too low to be 
well studied in a time-frequency analysis or by 
prewhitenning shorter parts of the light curve. We therefore do not 
consider them further in this work.

\begin{figure*}
\sidecaption
\includegraphics[width=12cm]{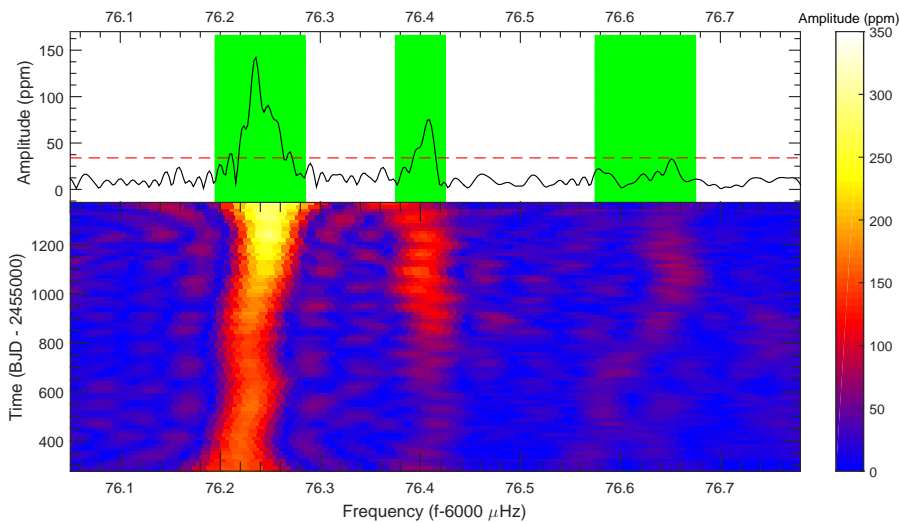}
\caption{Frequency and amplitude modulations of a group of linear 
combination frequencies $C_1$ near 6076\,$\mu$Hz. The 
red dashed line indicates the 5.6$\sigma$ detection threshold. 
The green shadowed areas in top panel represent the scales of 
variation of these frequencies (see text for details).
\label{lcf}}
\end{figure*}

Beyond the multiplets generated by the rotation of the star, we also 
focus on the interesting narrow frequency region near 6076\,$\mu$Hz 
where three structures ($f_{23}$, $f_{35}$ and $f_{74}$) 
show amplitude and frequency modulations as illustrated in Fig.\,\ref{lcf}.
These frequencies are in fact related to components of the 
$T_1$ and $T_2$ triplets through linear combinations. We find that
$f_{23}\sim{}f_1+f_{11}$, $f_{35}\sim{}f_3+f_{21}$, and 
$f_{74}\sim{}f_3+f_{11}$. Interestingly, we note that the frequency and 
amplitude variation pattern of $f_{23}$ is similar to the variations 
observed for the mode $f_1$. Similarities also exist between the variations 
observed in $f_{35}$ and $f_2$. The peak $f_{74}$, for its part, shows a 
rather large frequency variation covering up to $\sim0.1$\,$\mu$Hz 
(the scales of these frequency variations are indicated by the green 
shadowed region in the top panel of Fig.\,\ref{lcf}).

After this description of the various behaviors encountered, we concentrate, 
in the following sections, on plausible theoretical interpretations for 
the observed amplitude and frequency modulations.

\section{Resonant mode coupling and amplitude equations}

In this section, we recall the most natural theoretical background to 
understand the behavior of the modes forming the six multiplets induced by 
stellar rotation. These are indeed prone to develop nonlinear interactions 
through resonant mode couplings, which is the mechanism that we ultimately 
support from our present analysis. But before moving forward in discussing
details on the nonlinear resonant coupling mechanism, we first rule out 
several other possibilities as the cause of the observed modulations, such
as instrumental deffects, binarity, stochastically driven pulsations, or 
stellar "weather".

Instrumental modulations can possibly occur, e.g., on a per quarter basis, 
such as a slightly varying contamination from nearby stars that could 
affect the amplitude of the modes. Such effects would however affect all
frequencies similarly, which is not what is observed with KIC\,10139564 
where the modes show different types of behavior. Another effect related 
to the instrument that could induce frequency and amplitude modulations
is the slight shift of the Nyquist frequency associated with the movement 
of the {\sl Kepler} spacecraft in the Solar System barycentric reference 
frame. Fortunately, the multiplets that we consider (with frequencies 
below 5761\,$\mu$Hz), are far away from the Nyquist frequency limit 
($\sim 8496$\,$\mu$Hz). Moreover, such well-structured nearly equally spaced 
multiplets can obviously not be the mirror reflected frequencies of signals 
occurring above the Nyquist limit \citep{ba12}.

The presence of orbiting companions around compact stars could also induce 
frequency variations. However, these should occur in all frequencies and be
correlated in phase, such as in the sdB pulsator V391\,Peg \citep{si07}.
The variations that we find in several frequencies of KIC\,10139564 are 
clearly not correlated in phase. In addition, radial velocity measurements 
from spectroscopy do not show any significant variation, thus ruling out
the presence of a stellar companion \citep[][but for substellar objects, 
a higher precision would be needed to exclude this possibility]{ba12}.
 
Stochastically driven pulsations by envelope convection have long been 
observed in the Sun and solar-like stars. It has been claimed in the past
that stochastic oscillations could also occur in some sdB pulsators, based
on the observation that mode amplitudes could vary from season to season
\citep{ki10,re07}. \citet{os14} recently announced that stochastic pulsations 
were found in the sdB star KIC\,2991276,
in which the amplitude and phase of the modes vary substantially and 
irregularly on a timescale of a month. However, the mechanism responsible
for the oscillations in sdB stars, a well identified $\kappa$-effect 
involving iron-group elements \citep{ch97,fo03}, is very different in nature
from the stochastic driving occurring in the convective envelope of 
solar-like stars. Such a mechanism would indeed hardly be efficient in sdB stars 
that have radiative envelopes, except may be for a very narrow convective 
layer generated by the accumulation of iron in the Z-bump region (the latter 
being however extremely weak). 
Beyond these theoretical considerations, we find in 
the case of KIC\,10139564 that several mode behaviors, e.g., the 
frequencies in the triplet $T_1$ and the amplitudes in the doublet $D_1$, 
show correlations that would be difficult to account for with a stochastic 
driving mechanism and that essentially rule out this interpretation.

Changes in the background physical state of the star such as possibly 
induced by magnetic cycles could also be invoked for explaining amplitude and
frequency modulations. Magnetic cycles indeed have an impact on the frequencies of 
$p$-modes observed in the Sun and lead to small frequency drifts that 
correlate in time with tracers of the solar surface activity 
(see, e.g., \citealt{sa15} and references therein). However, there is no 
clear observational evidence of stellar activity on the surface of 
sdB stars which, again, have very stable radiative envelopes and are not 
known to be magnetic. Cycles comparable to those observed in solar-like 
stars are therefore unlikely to be found in sdB stars. Moreover, such a 
mechanism, or more generally a phenomenon modifying the physical state of 
the star on a timescale of months could hardly account for the observed 
modulations in KIC\,10139564 that show very different modulation behaviors 
from mode to mode, while a global change in the star would affect all modes 
similarly. Consequently, we also rule out this possibility in the present 
case.

\begin{figure}
\centering
\includegraphics[width=7.5cm]{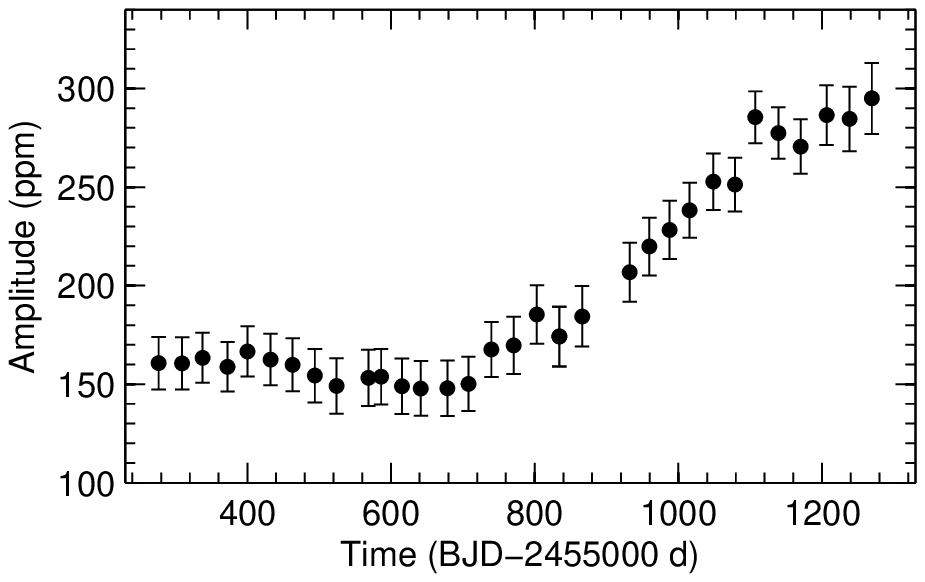}
\includegraphics[width=7.5cm]{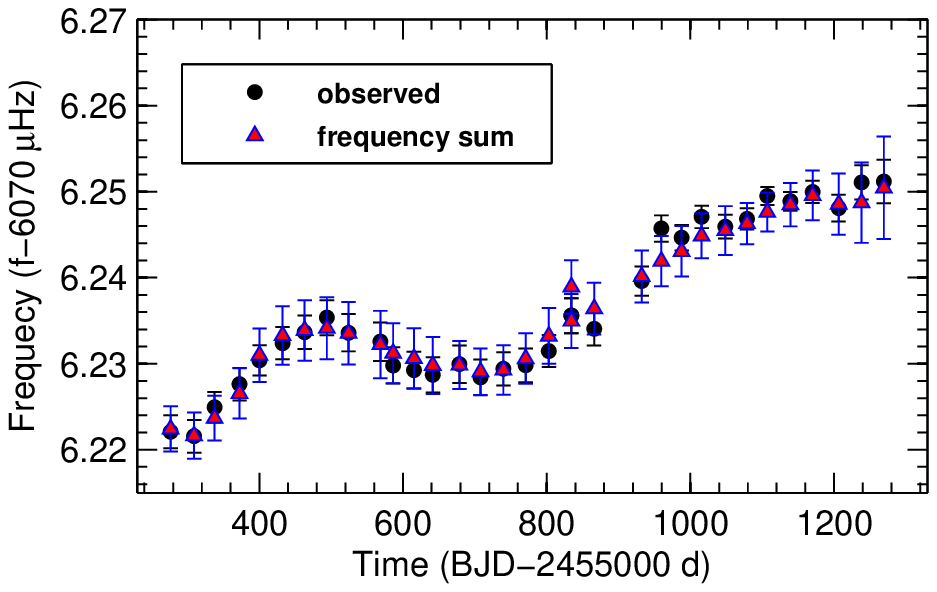}
\includegraphics[width=7.5cm]{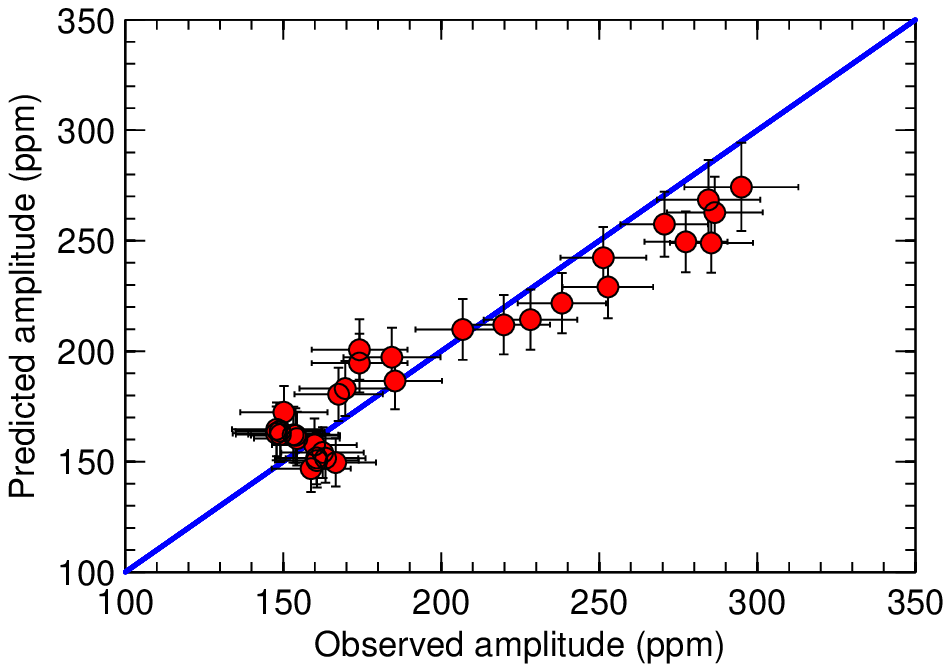}
\caption{Amplitude and frequency variations of the linear 
combination frequency $f_{23}=f_1+f_{11}$. 
{\sl Top panel}: Measured amplitudes as a function 
of time obtained from each data subset (using the same method as for 
multiplets). 
{\sl Middle panel}: Measured frequencies from each 
data subset compared with the frequency sum $f_1 + f_{11}$, both as 
a function of time.
{\sl Bottom panel}: Observed amplitudes of $f_{23}$ vs 
predicted amplitudes of $R \times$ the product of $f_1$ 
and $f_{11}$ amplitudes (see Eqn\,6 for the definition of $R$). 
In both cases (frequency and amplitude comparisons), the 
measurements are found to be within 1$\sigma$.
\label{lc1}}
\end{figure}

\begin{figure}
\centering
\includegraphics[width=7.5cm]{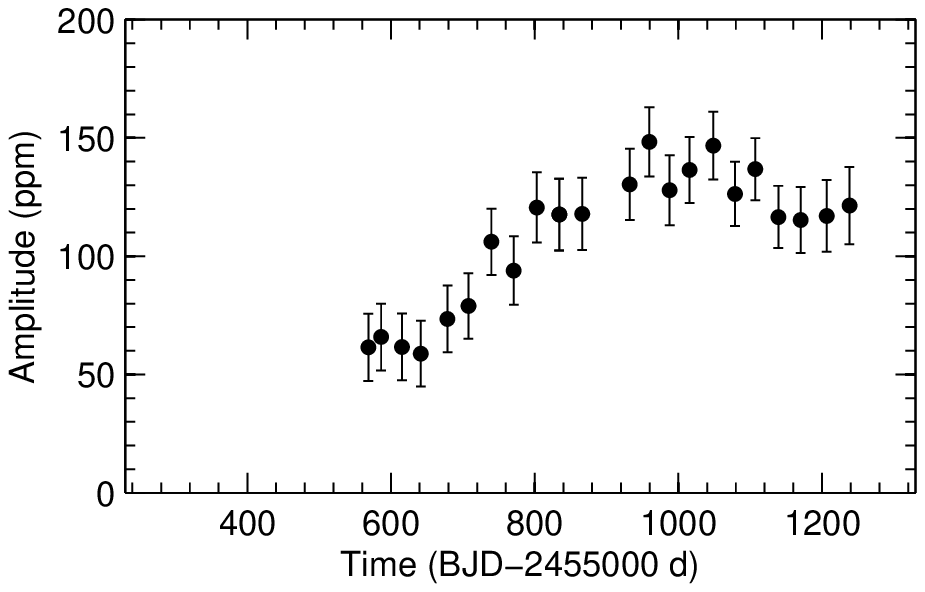}
\includegraphics[width=7.5cm]{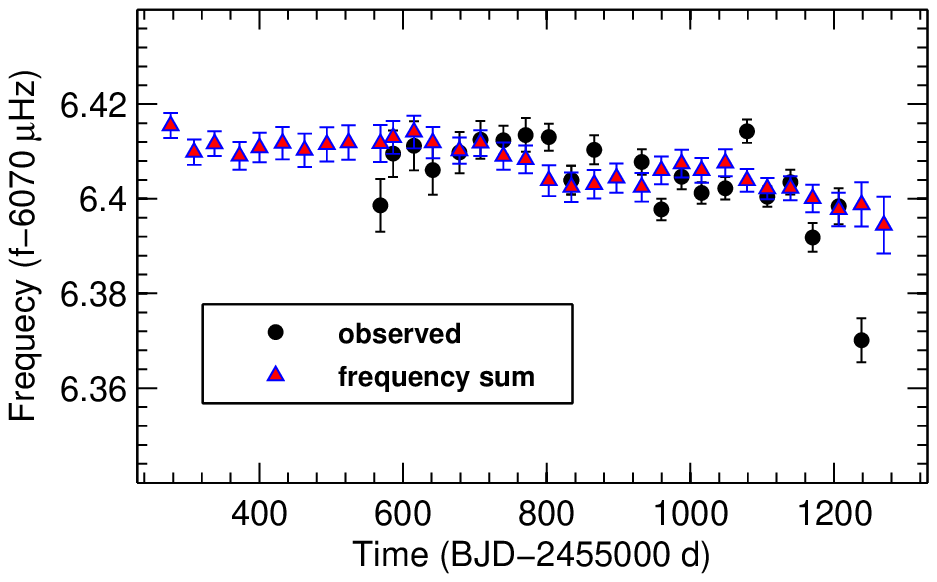}
\includegraphics[width=7.5cm]{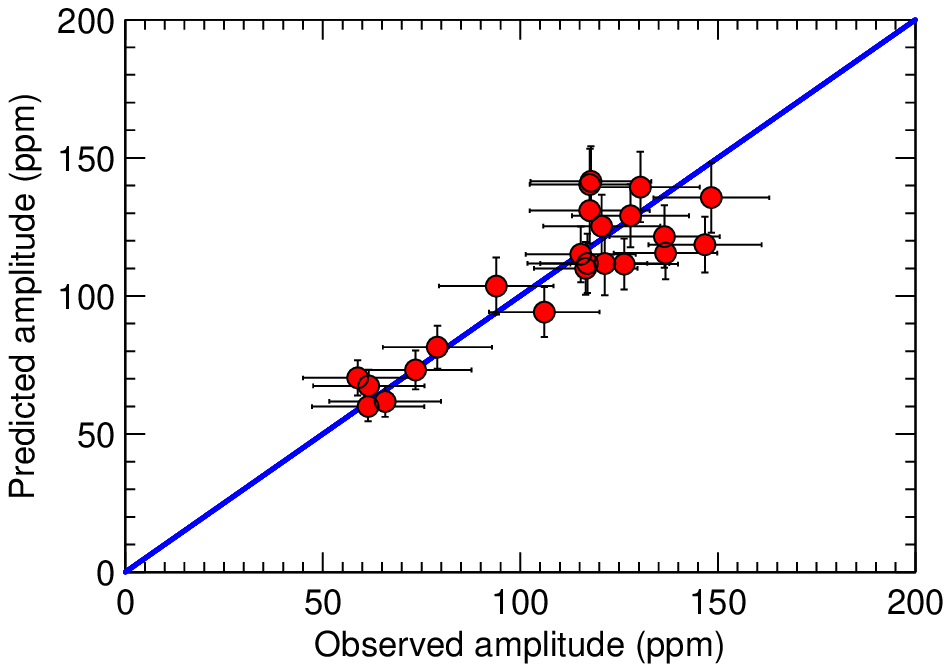}
\caption{Same as Fig.\,\ref{lc1} but for the linear combination 
frequency $f_{35}=f_3+f_{21}$. Note that there are 10 missing measurements 
for $f_{21}$, including the first 9 data points, because the 
amplitudes were lower than 4$\sigma$. The last data 
point is also not shown because of a large associated error, this measurement 
being at the end of the data set.
\label{lc2}}
\end{figure}

\subsection{Triplet resonance induced by slow stellar rotation}
We hereafter propose that nonlinear resonant coupling mechanisms could 
be a natural explanation for the observed modulations in KIC\,10139564.
Resonant interactions between modes may indeed result in amplitude and 
frequency variations occurring on timescales of weeks, months, and 
even years.

In the present context, we limit ourselves to the type of resonances 
described in \citet{bu95,bu97} involving linear frequency combinations 
$\nu_1\,+\nu_2\,\sim\,2\,\nu_0$. More specifically, we focus on a 
particular case where dipole ($\ell=1$) modes are split by slow rotation  
and form a nearly symmetric triplet (thus following the above relationship
between the frequencies of the components). We also consider the three-mode 
couplings of the form $\nu_1\,+\nu_2\,\sim\,\,\nu_0$, which corresponds
to the so-called direct resonances or parametric instabilities 
\citep{dz82,wu01b}.

We first recall some basic theoretical background relative to resonances in 
mode triplets created by stellar rotation. We emphasize that our 
focus on this particular mechanism is obviously motivated by the specific 
configuration of the modes observed in KIC\,10139564, most of which are 
identified as $\ell=1$ rotationally split triplets. We also point out that
this type of nonlinear resonance has recently been strongly suggested to 
explain the modulations of the {\sl g}-modes triplets in the DB white dwarf 
KIC\,08626021 (Z16). It is therefore the most natural effect that one could
think of in the present case.
The AEs formalism could also, in principle, be extended to multiplets of 
degree $\ell>1$ at the expense of solving a larger set of coupled amplitude 
equations \citep{bu95}. However, such development has yet to be done, 
which is beyond the scope of our present paper. The latter would be needed 
for KIC\,10139564 in order to fully interpret the several multiplets 
with $\ell>1$ that show variations.
The behavior for more complex $\ell>1$ multiplets may indeed differ from 
the simpler (better documented) $\ell=1$ triplet case, although we expect 
some similarities in general. 

Details on the theory of nonlinear resonant couplings for three-mode 
interactions, such as in $\ell=1$ triplet, can be found in 
\citep[][Z16]{bu95,bu97}. We summarize below the most relevant aspects 
(for our purposes) of the theory. In particular,
The quantity $\delta\nu$ (which we thereafter call the frequency asymmetry), 
measuring the departure from exact resonance (that would occur if, e.g., 
triplets were perfectly symmetric), is in fact essential for driving the 
various resonant mode coupling behaviors. 
Contributions to the frequency asymmetry in a given triplet generally 
involves higher order effects of stellar rotation on the pulsation 
frequencies \citep{dz92,jo89}, but could also have additional origins, 
such as the presence of a weak magnetic field\footnote{The asymmetry 
would be proportional to the strength of magnetic field 
$\lvert{}\vec{B}\lvert{}^2$ and the frequency of each component of the 
triplet (except the central, $m=0$, one) would be shifted in the 
same direction \citep{jo89}.}. We do not however consider further that 
possibility since no evidence of significant magnetism exists for 
sdB stars \citep{pe12,la12}.

The rotationally split frequencies up to the second order, which should be
the main contribution to the frequency asymmetry, are given by the formula
\begin{equation}
\nu_m-\nu_0 = (1-C_{k\ell})m\Omega + D_{k\ell}\frac{m^2\Omega^2}{\nu_0}, 
\end{equation}
where the $C_{k\ell}$ coefficient is the well-known first order Ledoux 
constant, $D_{k\ell}$ involves a complex integration of the eigenfunctions 
of the modes, and $\Omega\,=\,1/P_{\rm rot}$ is the rotation frequency of 
the star (expressed in Hertz).
The value of $C_{k\ell}$ is typically $\sim 0.5$ for dipole $g$-modes 
when approaching the asymptotic regime, while it is usually very small
($C_{k\ell} \ll 1$) for $p$-modes. The second order coefficient 
$D_{k\ell}$ is roughly 4$C_{k\ell}$ for dipole $g$-modes \citep{dz92,go98} 
but can vary significantly from one $p$-mode to another \citep{dz92,sa81}.  
The rotation period of KIC\,10139564, $P_{\rm rot}$, can be estimated from the
average of the frequency separations between the components of the multiplets
using the first order approximation $\Delta\nu=(1-C_{k\ell})\Omega$.
We find $P_{\rm rot}\sim 26$\,days for KIC\,10139564 (see Sect. 2). 
An "observed" frequency asymmetry can also be evaluated directly from the measured 
frequency of each triplet component, simply from the relation
\begin{equation}
\delta\nu_o = \nu_- + \nu_+ - 2 \nu_0.
\end{equation}
We note at this stage that $\delta\nu_o$ may actually differ from asymmetries expected 
from linear developments (such as discussed above) because nonlinear 
effects can modify the frequencies of the modes.

The numerical solutions of the AEs for mode interactions in triplets 
mainly reveal three distinct regimes of resonances (see, e.g., 
\citealp[][and in Z16]{bu97}). The first state is the "frequency lock" 
regime where all the components in the triplet have constant frequencies 
and amplitudes and the asymmetry tend to be zero (triplets become perfectly 
symmetric). The opposite configuration is the nonresonant regime 
where the triplet configuration is like predicted by the linear theory of 
stellar oscillations. Between the two, there is an intermediate regime in 
which all the modes in the triplet show modulated frequencies and 
amplitudes which can be periodic, irregular, or even chaotic.

In order of magnitude, the occurrence of these three regimes is roughly 
linked to a parameter $D$ defined as 
\citep[see][]{go98}
\begin{equation}
D \equiv \frac{2\pi\delta\nu}{\kappa_0},
\end{equation}
where $\kappa_0$ is the linear growth rate of the $m=0$ mode in the triplet 
(a nonadiabatic quantity). However, the ranges of values for this parameter 
which define the different regimes depend somewhat on the values of the 
nonlinear coefficients in the real star. $D$ is also a quantitative indicator 
that measures how far the triplet modes are away from the exact resonance 
center ($D=0$). We nonetheless summarize some of the properties encountered
in previous studies as a function of $D$ :

\begin{itemize}
\item In the frequency lock regime ($\delta\nu \rightarrow 0$),   
the $D$-parameter roughly corresponds to values in the range $\sim 0-1$ 
according to the AEs formalism. However, in the case of the white dwarf 
star GD358, \citet{go98} found that $D$ could be up to 20 and still 
correspond to a frequency locked situation. These ranges, therefore,
are somewhat dependent on the specific properties of the mode being 
considered, in particular on the scale of their linear growth rate, 
$\kappa_0$.\\

\item The intermediate regime occurs when the triplet components move away 
from the resonance center ($\delta\nu\ne0$). In this situation, if periodic 
variations indeed affect the considered modes, these can be 
expected to have a modulation timescale of 
\begin{equation}
P_{\rm mod} \sim \frac{1}{\delta\nu}\simeq\frac{2\pi}{\kappa_0}\frac{1}{D},
\label{eqnpmod}
\end{equation}
i.e., roughly the timescale derived from the inverse of the linear (i.e., 
unperturbed) frequency asymmetry of the triplet dominated by the second 
order effect of stellar rotation (following Eqn.\,2). This timescale is 
also connected to the inverse of the growth rate of the oscillation mode 
through the $D$ parameter \citep{go98}.\\

\item The modes recover a configuration of steady pulsations with the 
nonresonant regime when the involved frequencies are such that the modes 
are now far from the resonance condition ($D >> 1$). In this regime, the 
nonlinear interaction between modes is very weak and nonlinear frequency 
shifts become very small. Consequently, the mode frequencies are close to 
the linear ones.
\end{itemize}

We finally point out that in addition to the above mentioned three main regimes, 
a narrow hysteresis (transitory) regime exists between the frequency lock 
and intermediate regimes in which the frequencies can be locked (i.e., constant), 
while the amplitudes still have a modulated behavior \citep{bu97}.

\subsection{Three mode resonance of the type \threemodes{}}
In this section, we recall some properties of nonlinear interactions 
between oscillations modes not within triplets but whose frequencies are 
close to a resonance condition such that \threemodes{}. 
Frequencies with such a relationship could also result from
simple linear combination frequencies, i.e., exact sum or difference of 
frequencies (where the "child" frequency is not a true eigenmode), 
which may be related to nonlinearities in the mixing process affecting the depth
of a convective zone in the outer layer of a pulsating star \citep{wu01a}, or 
to nonlinearities in the flux response induced by the surface geometrical and 
temperature distorsions triggered by the propagating waves \citep{br95}. 

A useful quantity, $R$, connecting the observed amplitude of the combination 
frequency and the amplitudes of its "parent" modes, has been defined as 
\citep[][]{va00,wu01a} 
\begin{equation}
R = \frac{A_0}{A_1\cdot{}A_2},
\end{equation} 
where the $A_0$, $A_1$, and $A_2$ are the amplitudes of the frequencies 
$\nu_0$, $\nu_1$, and $\nu_2$, respectively.
This ratio $R$ is typically less than 10 for simple linear combinations 
related, e.g., to nonlinearities in the flux response. Consequently, 
in pulsating sdB stars, the "child" frequency resulting from this effect 
usually has a very low amplitude compared to its "parent" 
frequencies. In the large amplitude and brightest known 
pulsating sdB star, Balloon\,090100001, where such linear combination frequencies 
have been unambiguously observed, the amplitude ratios are 3.9, 3.7, 3.0 and 5.5 
for the linear combination frequencies of four {\sl p}-modes and one 
{\sl g}-mode $f_1+f_2$, $f_1+f_3$, $f_1+f_4$ and $f_1-f_B$ in 
$B$-band photometry \citep{ba08}, respectively. In the 
present work, we however find that the identified linear combination 
frequencies $C_1$ have amplitude ratios in the $10-100$ range, i.e., 
one order of magnitude larger than typical linear combination frequencies
observed so far. {One possible interpretation for the high amplitude ratios 
is that the frequency sum/difference is near the resonance condition of 
\threemodes{} and its amplitude is boosted significantly by the resonance 
\citep[e.g.,][]{dz82,br14}}

The AEs formalism treating the \threemodes{} type of resonance, including 
the parametric instability and the direct resonance (see below), 
is similar to the case of a triplet resonance 
\citep[e.g., see the amplitude equations in][Z16]{bu97}. According to \citet{dz82}, 
the three-mode interactions can be described by the following coupled system 
\begin{subequations}
\begin{align}
  \frac{d\vec{A_0}}{dt} & = \kappa_0\vec{A_0} + i\frac{q}{2\nu_0 I_0}\vec{A_1A_2}\exp(-i\delta\nu{}t),  \label{seq61} \\              
  \frac{\vec{dA_1}}{dt} & = \kappa_1\vec{A_1} + i\frac{q}{2\nu_1 I_1}\vec{A_0A_2^*}\exp(-i\delta\nu{}t), \label{seq62} \\
  \frac{d\vec{A_2}}{dt} & = \kappa_2\vec{A_2} + i\frac{q}{2\nu_2 I_2}\vec{A_0A_1^*}\exp(-i\delta\nu{}t). \label{seq63}   
\end{align}
\end{subequations}
Where $\vec{A_j^*}$ is the complex conjugate of the amplitude $\vec{A_j}$
($\vec{A_j}=A_je^{i\phi_j}$),
$I_{j}$ is the mode inertia, and $\kappa_{j}$ is the linear growth rate 
for the three involved frequencies. The quantity $q$ is a nonlinear 
coupling coefficient, and $\delta{}\nu$ is the frequency mismatch relative to 
pure resonance defined by the relationship $\delta{}\nu=\nu_0-\nu_1-\nu_2$.

The nonlinear equations (\ref{seq61}), (\ref{seq62}) and (\ref{seq63}) cannot be 
solved by analytical methods, but solutions for the equilibrium state (all time derivatives 
set to zero) can be obtained. In particular, the equilibrium solution leads to an amplitude 
ratio 
\begin{equation}
R = \frac{q}{2\nu_0\kappa_0I_0} \equiv \frac{A_0}{A_1\cdot{}A_2}.\label{nonratio}
\end{equation}
{The stability of the equilibrium-state depends on the growth (damping) rates and the 
frequency mismatch \citep[e.g.,][]{dz82}.}

In a three-mode direct resonance, the child mode is damped and has a frequency
very close to the sum of frequencies of its two parent modes which are linearly driven 
(unstable). The child mode amplitude is very sensitive to its mode inertia, linear 
growth rate, and to the nonlinear coupling coefficient (see Eqn.\,\ref{nonratio}). 
This near resonance mode can grow up to a very large amplitude if the 
quantity $q/\kappa_0{}I_0$ is sufficiently large. 
The coupling coefficient $q$ follows from a complex integration 
of the coupled mode eigenvectors and its explicit form can be found in 
\citet{dz82}. It may be possible, in principle, to calculate this 
coefficient provided that the mode eigenfunctions are known. However,
this would require that a precise seismic solution is found for KIC\,10139564, 
which still has to be obtained.
To test the theory of a three-mode direct resonance, we would also need to 
know the linear damping (growth) rate and the inertia of the damped 
mode, eventually corrected by the effect of slow stellar rotation 
\citep[e.g, see][]{ca82}. {However, the situation could be simplified 
in the case of unstable equilibrium state where the amplitude and 
frequency of the child mode should exactly follow those of its parent modes, 
even if the growth rates and coupling coefficients are unknown. Fortunately, 
the equilibrium state of three-mode direct resonances seems always unstable 
because the growth (damping) rates cannot satisfy the stability criteria 
of the Hurwitz theorem \citep{dz82}. Therefore, each frequency and amplitude
measurement could be used as one independent test of these particular
nonlinear couplings. Furthermore, this also provides a method to 
separate the child mode from their parent modes according to the amplitude 
and frequency relationships \citep[e.g.,][]{br14}.}

The parametric instability is another form of three-mode resonant coupling 
that could destabilize a pair of stable daughter modes from an 
overstable (driven) parent mode \citep{wu01b}. 
In this mechanism, the overstable parent mode gains energy through the
driving engine (a $\kappa$-mechanism in our case) and the two independent 
damped child modes dissipate energy. This configuration would lead 
the system to reach limit cycles under certain conditions 
\citep[e.g., if $\delta\nu<\kappa_1$ or $\kappa_2$, ][]{wu01b,mo85}. 
During such limit cycles, the amplitude of the parent mode first
increases slowly on a timescale of $\kappa_0^{-1}$, then decreases 
rapidly on a timescale of $\kappa_{1,2}^{-1}$. A the same time, the amplitude 
of the daughter modes follow the opposite behavior. 
In sdB stars, we point out that the linear growth rate of the parent 
mode $\kappa_0$ would be usually far smaller than the damping rate of 
the daughter modes $\kappa_{1,2}$.
We further mention that the nonlinear interactions between the parent/child modes 
would also affect their periods as a result of phase variations. 
The nonlinear frequency shift could be of the order of a few $\mu$Hz 
in some extreme conditions \citep{wu01b,mo85}. We point out that a 
parametric instability can also occur in multiplets. In such 
circumstances, different $m$ components that forms the multiplet may share 
some common damped daughter modes. Having common daughter modes 
involved in different parametric resonances, i.e., involving 
different parent overstable components, will obviously induce more
complex dynamic modulations than simple periodic variations that could be 
expected from pure three-mode only interactions.
We indeed point out that both the triplet resonance that was 
explored by \citet{bu95,bu97} and the three-mode 
\threemodes{} resonances that were investigated by \citet{dz82}, 
\citet{mo85} and \citet{wu01b} are treated as isolated systems, 
i.e., assuming only interactions between the three involved modes and 
ignoring the possible influence of other modes. Modes with the 
highest amplitudes are more likely to efficently couple with different 
resonances, e.g., in a multiplet resonance and in a \threemodes{} 
resonance.

\section{Connections with mode behaviors seen in KIC\;10139564}

In light of the theoretical background summarized in the last section, 
we tentatively interpret some of the behaviors described in Section 2 for 
the frequencies listed in Table\,1. These indeed show striking similarities 
with patterns expected for nonlinear resonant mode interactions that occur 
in various regimes.

\subsection{Multiplets in the intermediate regime}

The first connection is for the modes belonging to multiplets that 
show quasi-periodic amplitude and frequency modulations. In particular, the 
$p$-mode triplet near 5760\,$\mu$Hz ($T_1$) shows indication that it could be evolving
within the so-called intermediate regime of a triplet resonance. We recall 
(see Section 2;  Fig.\ref{pr1}) that the frequencies of the two side components in 
this triplet, besides showing a long term drift, vary quasi-periodically in 
antiphase with a timescale of $\sim550$ days. The central component 
of $T_1$, for its part, has a frequency modulation which also vary, possibly with 
a slightly longer period of $\sim600$ days. 
For comparison purposes, the modulation timescale is expected to be related to 
the inverse of the linear (i.e., unperturbed) frequency asymmetry in the 
triplet (see Eqn.\ref{eqnpmod}), which therefore should be $\delta\nu\sim 0.02$ 
$\mu$Hz. Assuming that this frequency asymmetry originaly comes from the second 
order effect of slow rotation, and given the average rotation frequency of 
the star ($\sim 0.42$ $\mu$Hz, corresponding to $\sim 26$ days), the $D_{k\ell}$
coefficient in Eqn.~1 can be estimated to $\sim 200$ for that mode. This value is
plausible because the $D_{k\ell}$ coefficient is found to vary over a large range
for dipole $p$-modes \citep{dz92,sa81}. However, to compute $D_{kl}$ 
in this specific case and compare with this value, a precise seismic solution
for KIC\,1013956 has to be worked out, but is not available yet.
It has to be noted that the frequency asymmetry measured from the averaged
frequencies given in Table\,1 is only 0.0026\,$\mu$Hz that is one order of 
magnitude lower than the value derived from the modulation frequency 
($\sim0.02$\,$\mu$Hz). We note, however, since the frequencies are varying with time,
that the maximum extent of the observed frequency asymmetry is $\sim0.02$\,$\mu$Hz 
when considering the 33 measurements independently (see Fig\,\ref{p1}).
We point out that these observed values (0.0026\,$\mu$Hz on average and 
$\sim0.02$\,$\mu$Hz for the maximum asymmetry) are similar to those observed 
in the main triplet of \dbv{}, which is also in the intermediate regime (Z16).
Nonlinear resonant interactions are bound to perturb the linear 
frequencies of the modes, forcing them in some cases to shift toward the 
exact resonance (obtained when the system is locked). It is therefore not 
surprising to observe a frequency asymmetry that can be significantly smaller 
than the theoretical shift expected in the linear theory context. 

In terms of amplitude modulations, the situation is bit less clear as only the 
prograde component in $T_1$ shows a quasi-periodic modulation, with a 
timescale of $\sim800$\,days, while, for the other two components, particularly 
for the retrograde mode, their amplitude variations appear somewhat 
irregular. 

In addition to the frequency variations of $T_1$ discussed above,
we note that the three components that form this triplet feature a 
regular drift toward each other which, if nothing change, would lead
them to merge into one frequency on a timescale of $\sim10$ years.
Such a merging is of course not conceivable and what we observe is more
likely a small fraction of a variation cycle occurring on a timescale 
much longer than the duration of the {\sl Kepler} observations.
This suggests that the triplet resonance is probably not the only 
mechanism that affects the stability of $T_1$. This added complexity 
may also explain the more erratic behavior of the amplitude variations 
in this triplet. Quite notably, we indeed find that all the components 
of $T_1$ can be linked to other frequencies forming linear 
combinations satisfying the conditions for a three-mode 
resonance \threemodes{}. This will be further discussed in Sect.\,4.4.

The quintuplet $Q_1$ also shows components with amplitude 
and frequency variations (see Fig.\,\ref{q1}) that may be
associated to the intermediate regime. 
In this case, however, we cannot estimate timescales for the 
modulations which appear to have a more complex behavior
than the modulations detected in the $T_1$ triplet or, if 
we compare to other cases, in the triplets of the pulsating 
DB star KIC\,08626021 (Z16).
The averaged frequency mismatch, $\delta\nu_o$, for $Q_1$ is 
about 0.0018\,$\mu$Hz. This could either be the result of 
the nonlinear coupling mechanism locking the modulated 
components close to the exact resonance, even if they are 
in the intermediate regime (see the case of $T_1$, as well
as the triplets in the DBV star KIC\,08626021), or it could 
indicate that the modulation timescale for $Q_1$ is 
$\sim 17.6$ yrs (the inverse to $\delta\nu_o$, 
{if their amplitudes have a periodic behavior}). 
As there has been no theoretical exploration of the nonlinear 
five-mode interaction yet, the connection of $Q_1$ with the 
intermediate regime is based on the assumption that nonlinear 
five-mode interactions has also mainly three distinct regimes. 
The coupled amplitude equations for the five-mode resonance 
involve more terms in each AE and the numerical solutions 
are more difficult to search for.  

Another case may be connected with the intermediate regime:
the multiplet $M_1$ which shows amplitude and frequency 
variations (see Fig.\,\ref{s1}). But, again, we cannot 
determine any timescale for the complex modulations 
occurring in this multiplet. In that case, there is also a 
slow trend leading frequencies, particularly for the most side 
components, to seemingly converge. This trend is very similar 
to the slow variation observed in the $T_1$ triplet. It could
possibly be a fraction of a variation cycle with a much 
longer timescale than the duration of the observations, but 
more observations would be needed to confirm this hypothesis.
This multiplet $M_1$ should be the siege of even more complex 
resonant coupling interactions than the quintuplet $Q_1$, since 
there are six detected components, with at least three 
components missing.

\subsection{Triplets in the transitory regime}

Another type of behavior encountered in our data can be linked to
the narrow transitory hysteresis regime which is between the frequency 
lock and intermediate regimes. This state is characterized by stable 
frequencies but varying amplitudes. This is notably observed in the 
$g$-mode triplet $T_2$ (see Fig.\,\ref{g1}). For this triplet,
the observed frequency mismatch is about 0.0036\,$\mu$Hz, 
i.e., very similar to the value measured for the $T_1$ triplet 
(see left-top panel of Fig.\,\ref{p1} and \ref{g1}).
We also point out that $T_2$ may couple with the $p$-mode 
triplet $T_1$ through a three-mode resonance \threemodes{}, 
as discussed in Sect.4.4.

The incomplete triplet $D_1$ may also be associated to this transitory 
regime as it shows quasi-periodic amplitude modulations and 
stable frequencies. Due to the missing component, we cannot measure 
the frequency mismatch for this doublet. We note that the AEs
for the triplet resonance indicate that the modes cannot be stable, 
i.e, there is no fixed-point solution, if one of the visible modes 
forming the incomplete triplet is the central ($m=0$) component \citep{bu95}.
Thus, at this stage, we may just fail to detect either the 
third component of the triplet whose amplitude may be lower than
the detection threshold (meaning that the triplet is indeed in the 
narrow transitory regime), or the nonlinear modulation of the frequencies,
which may be smaller in amplitude than our current precision (meaning a 
doublet in the intermediate regime, as predicted by the AEs).

\subsection{A triplet in the frequency lock regime}

The last case occurring in a different regime is the $g$-mode triplet 
$T_3$, which shows stable amplitudes and frequencies (see Fig.\,\ref{g3} 
and \ref{sg3}). This suggests that $T_3$ is in the configuration of the 
frequency lock regime where the triplet approaches the resonance 
center, i.e., $\delta\nu\rightarrow 0$ and both the frequencies and 
amplitudes are constant. Indeed, we find that the observed frequency 
asymmetry, $\delta\nu_o$, is 0.0008\,$\mu$Hz (Fig.\,\ref{g3}) for 
$T_3$, which is less than the measured error 0.0011\,$\mu$Hz. The triplet 
$T_3$ is therefore exactly (within measurement errors) at the resonance 
center, contrary to $T_1$ and also $T_2$ which has constant frequencies 
but a small non-zero frequency mismatch.

In summary, the various behaviors encountered in the multiplets detected
in KIC\,10139564 seem to cover all the different regimes expected in a 
context of resonant mode coupling in multiplets. This mechanism is 
therefore quite likely responsible, at least in part, for the observed 
phenomena. In the following section, we discuss another type of 
resonance, the \threemodes{} nonlinear interaction.

\subsection{Three-mode resonance}

In this section, we discuss the variations of a group of 
frequencies $C_1$, including $f_{23}$, $f_{35}$ and $f_{74}$, 
that are involved in a relationship \threemodes{}. 
We find that the variations of these frequencies have 
strong correlations with the variations of the components 
in the triplet $T_1$ (see Fig.\,\ref{p1} and \ref{lcf}).
The large frequency variations first suggest that the $C_1$ 
frequencies correspond to three-mode resonances rather than 
simple linear combination frequencies.
The result of prewhitening the frequencies  $f_{23}$ and $f_{35}$
(using the same method as for the multiplets) is shown in 
Fig.\,\ref{lc1} and \ref{lc2}, respectively. 
Most of the amplitude and frequency measurements for $f_{23}$ and $f_{35}$
are exactly following the variation of amplitude and frequency of 
the sums $f_1+f_{11}$ and $f_3+f_{21}$ within 1$\sigma$, respectively 
(see in particular the middle and bottom panels of Fig.\,\ref{lc1} and 
\ref{lc2}).

The amplitude ratio $R$ is 37 and 85 for $f_{23}\sim f_1+f_{11}$ 
and $f_{35}\sim f_3+f_{21}$, respectively. These values are significantly
higher than those observed for normal linear combination frequencies 
in sdB stars (see the example given in Sect.\,3.2). There is also a possible 
true linear combination frequency in KIC\,10139564 with the frequency 
$f_{79}\sim f_1$\,-\,$f_4$ (see Table~2) which indeed has a very low 
amplitude (signal-to-noise ratio of 5.1) and an amplitude ratio $R$ less 
than one. Thus, we propose that there should be real pulsation modes near 
the position of the linear combination frequencies $f_1+f_{11}$ and 
$f_3+f_{21}$ and these modes had their amplitudes boosted through a 
resonance. In the \threemodes{} resonance, the child mode indeed 
follows the behavior of its parent modes (see, again, Fig.\,\ref{lc1} 
and \ref{lc2} and examples provided by \citealt{br14}).

We note that another frequency, $f_{74}$, is also in the
region near 6076\,$\mu$Hz, but has an amplitude too low to be monitored
over time using the prewhitening technique on subsets of the data.
However, Fig.\,\ref{lcf} still clearly shows that this frequency is 
varying smoothly during the observation, from $\sim6076.58\,\mu$Hz 
(the first half part of the run) to $\sim6076.68\,\mu$Hz (the second
half part). We speculate that there is possibly a real mode, with a 
frequency around $6076.58-6076.69\,\mu$Hz, which first interacts with 
the frequency sum $f_2+f_{39}\sim6076.59\,\mu$Hz, then with the 
frequency sum $f_3+f_{11}\sim6076.66\,\mu$Hz, because the influence 
from the latter modes become stronger than the former ones during the last 
half of the observation run, due to the amplitude of $f_3$ increasing 
significantly in the second half of the {\sl Kepler} time series. 
This, again, suggests that the $C_1$ frequencies are really part of 
\threemodes{} resonances instead of being simple linear combination 
frequencies.

All of the involved frequencies, $f_{1, 2, 3, 11, 21, 39}$, 
are the components of the triplets $T_1$ and $T_2$. They are expected 
to be overstable (driven) modes, thus meaning that they are involved in
a three-mode direct resonance and not a parametric resonance which 
involves one overstable parent mode and two damped unstable daughter modes.

At this stage, it becomes natural to interpret the complex variations 
observed in the components of $T_1$ and $T_2$ to be linked with the fact 
that these modes are simultaneously involved in two different types of 
resonances, i.e., a triplet resonance and \threemodes{} direct resonances.
In this situation, the triplet resonance may be the dominating nonlinear 
interaction occurring in the triplet, while the nonlinear coupling with 
the modes outside the triplet could strongly perturb the periodic 
amplitude and frequency modulations expected if the triplets were pure 
isolated systems. This shows that nonlinear mode interactions in real 
stars are certainly more complex configurations than those treated by 
current simplified theoretical approaches.
Moreover, since the $T_2$ triplet is in the transitory regime, 
with frequencies locked by the nonlinear coupling within the triplet, 
the nonlinear interactions outside this triplet are therefore unable 
to destroy this frequency locking, resulting in no long-term frequency 
variation as can be seen in the $T_1$ triplet.
Interestingly, the resonant mode coupling theory predicts that a limit 
cycle (steady equilibrium state) may not be reached in the case of 
three-mode direct resonance (which is likely at work here as discussed 
above). The evolution of this long-term frequency variation in $T_1$, 
whether the mode frequencies will further converge or eventually 
diverge, and the evolution of $T_2$, whether frequencies will remain 
constant or the locked regime will eventually be broken, will need further 
observation either from ground or by future space instruments currently in 
preparation, e.g., TESS and PLATO \citep{ra14,ri14}.

\subsection{The $D$-parameter and further insight on the modulations}

The value of the $D$-parameter, that defines how far the modes are from the 
resonance center, is usually connected to the kind of regime a multiplet 
is in when undergoing resonant mode interactions.
This $D$-value is in particular sensitive to the linear growth rate of 
the oscillation modes (see again Section 3). 

For the frequency lock regime that is observed in KIC\,10139564 with 
the $T_3$ triplet, $D$ is near or exactly zero, as predicted by the 
AEs \citep{bu95,go98}. 
The $D$-value for the other triplets $T_1$ ($\delta\nu_o \sim 0.0026$\,$\mu$Hz) 
and $T_2$ ($\delta\nu_o \sim 0.0036$\,$\mu$Hz) may reflects more their linear 
growth rates since, with very similar frequency mismatches, the two triplets are 
found to be in different regimes. The growth rate values are indeed 
substantially different between $p$- and $g$-modes \citep{ch99,fo03}. 
Assuming the growth rate for $p$-modes is of the order of 10$^{-6}$\,s$^{-1}$ 
\citep{ch99}, the corresponding $D$ value for the $T_1$ triplet would be far less 
than 1, indicating that $T_1$ should be in the frequency lock regime 
\citep{bu97,go98}, but we find it to be in the intermediate regime.
We note however that this estimate for the value of $D$ is based on the measured 
frequency mismatch which may not be representative of the unperturbed frequency
asymmetry that enters in the definition of $D$. The latter is likely 
much larger (see Section 4.1), leading to a somewhat larger $D$-value more
in line with the observed regime of the resonance for $T_1$.
The $D$-value for the $g$-mode triplet $T_2$, for its part, could be much 
larger than that of $T_1$, considering the much smaller growth rates of 
the $\ell=1$ $g$-modes \citep{fo03}. Extended ranges for $D$ were also 
found in \dbv{} (Z16), where the $D$-values for the triplets are at least 
two orders of magnitude larger than those suggested in \citet[]{go98}.
This suggests that the nonlinear behaviors not only depend on the magnitude of 
$D$, but also on the specific coupling coefficients for each specific mode
\citep{bu95}.  

Further quantitative comparisons between the observed modulations and 
the theoretical framework would require to solve the amplitude equations 
for the specific case of KIC\,10139564. This would require to calculate 
the coupling coefficients in the AEs, which, in principle, could be 
extracted from the observed amplitude and frequency modulations 
\citep{bu95}. With these known coupling coefficients, one could then 
determine the ranges of $D$-values related to each different regime 
of the nonlinear resonance. A measurement of the growth rates of the 
oscillation modes would then possibly follow with the determination of this 
parameter, which may lead for the first time to an independent estimation of 
the linear nonadiabatic growth rates of the modes and a direct test of 
nonadiabatic pulsation calculations in sdB stars.

\begin{balance}

\section{Summary and conclusion}

While studying the high-quality and long-duration photometric data provided 
by the {\sl Kepler} spacecraft on the pulsating sdB star KIC\,10139564,
we have identified different patterns in the frequency and amplitude 
modulations of the oscillation modes belonging to several rotationally 
split multiplets or linear combination frequencies.
These modulations show signatures that can be associated to nonlinear 
resonant mode coupling mechanisms that could occur between the
multiplet components themselves and with other modes under certain 
conditions, i.e., satisfying a \threemodes{} resonance relationship. 
This is the first time that such signatures are quite clearly identified 
in pulsating hot B subdwarf stars, and the second case reported so far for 
a compact pulsator monitored with {\sl Kepler} photometry (see Z16).

We first reanalysed the 38-month of {\sl Kepler} photometry obtained 
for KIC\,10139564, leading to the detection of 60 independent 
frequencies above a secured detection threshold (5.6$\sigma$; see 
Table\,\ref{t2}). 
Among these, 29 frequencies consist of three triplets, one doublet, 
one quintuplet and two incomplete multiplets with $\ell>2$ 
(see Table\,\ref{t1}). Another three detected 
frequencies are linked to other frequencies through linear combinations. 
Five additional groups of frequencies are found in the region 
between 5400 and 6400\,$\mu$Hz, which have very complicated 
structures. 
Finaly, we also find 14 independent frequencies and two 
frequencies satisfying linear combination relationships that could 
be real as their amplitudes are between 5 and 5.6$\sigma$ above the
noise. In general, our well secured frequencies are in good agreement 
with the former analysis from \citet{ba12}. In this paper, we particularly 
concentrated our study on six multiplets and three linear combination 
frequencies observed near 6076\,$\mu$Hz.

We found different types of mode behaviors occurring in the 
above mentioned frequencies. A "short timescale" quasi-periodic amplitude 
and frequency modulations along with a slow trend of the frequencies to 
convergence toward each other occur in the dominant $p$-mode triplet near 
5760\,$\mu$Hz ($T_1$). The $\sim570$-day quasi-periodic frequency 
modulation evolve in antiphase between the two side components in this
triplet. Modulated frequencies and amplitudes are also found in 
a quintuplet near 5287\,$\mu$Hz ($Q_1$) and a ($\ell>2$) multiplet near 
5412\,$\mu$Hz ($M_1$), but the modulations do not show a clear periodicity. 
One triplet near 316\,$\mu$Hz ($T_2$) has a distinct 
behavior from the above mentioned multiplets, as it shows stable frequencies
but varying amplitudes. A similar phenomenon occurs in a doublet 
near 394\,$\mu$Hz ($D_1$) which shows constant frequencies and 
an $\sim 1100$ days periodic amplitude modulations. Another triplet at 
518\,$\mu$Hz ($T_3$) completely differs from all the above multiplets, 
with constant amplitudes and frequencies throughout the whole observation run. 
In addition, we also discovered amplitude and frequency variations in three 
frequencies near 6076\,$\mu$Hz ($C_1$) that are linked to other independent 
frequencies through linear combinations.

After ruling out various possible causes for the modulations,
we showed that these mode behaviors could be related to the 
different types of nonlinear resonances that should occur 
according to the amplitude equation formalism. In particular, 
nonlinear resonant couplings within a multiplet can lead to three 
main regimes, all of which are possibly occurring in  KIC\,10139564.
The multiplets $T_1$, $Q_1$ and $M_1$ can be associated 
with the intermediate regime of the resonance where 
the involved modes have modulated amplitudes and frequencies. 
The triplet $T_2$ and doublet $D_1$ have a different behavior that 
could be associated to a narrow transitory regime in which the 
frequencies of the modes can be locked (constant) while the amplitudes 
experience modulations. The behavior of the triplet $T_3$ is the unique 
case found in this star that can be associated to the frequency 
lock regime of the resonance, where both amplitudes and 
frequencies are stable. In addition, the large amplitude ratios 
between the $C_1$ frequencies and their main parent modes, 
together with the large variation of amplitude and frequency 
observed for these peaks, suggest that $C_1$ correspond to three-mode 
direct resonances. We indeed found that the frequencies of $C_1$ 
exactly follow the evolution of their main parent modes. Moreover, 
as the parent modes of $C_1$ are also the components of $T_1$ and 
$T_2$, we suggest part of the complexity of the mode behaviors could be
related to these cross interactions between the various modes. In particular,
the slow variations occurring in $T_1$ may be related to the \threemodes{} 
resonance superimposed to the triplet resonance occurring between the 
components.

We emphasize that the observed frequency modulations likely induced by 
nonlinear mode interactions could challenge future attempts 
to measure the evolutionary effects on the oscillation mode periods 
in pulsating sdB stars. Compared to the resonant variations taking place 
on timescales of years, the rate of period change of the pulsations due 
to stellar evolution in sdB stars is much longer, typically occurring on 
a timescale of $\sim 10^6$\,yr \citep{ch02}. 
Nonlinear modulations of the frequencies can potentially jeopardize any 
attempt to measure reliably such rates, unless they can be corrected 
beforehand.
These nonlinear modulations could also complicate the detection of 
exoplanets or stellar companions around sdB stars using the 
technique of measuring phase changes in the pulsations \citep{si07}. 
It should be possible however to distinguish between the two effects, 
considering that nonlinear couplings may induce different behaviors 
on different modes, while external causes such as an orbiting body 
should affect all modes similarly.

Finally, we note that our analysis suggests that resonances 
occurring in real stars, in which modes could be involved in two 
or more types of different couplings, lead to more complicated 
patterns than those predicted by current theoretical frameworks which 
treat the modes only as isolated systems within one type of resonance 
and ignore the nonlinear interactions that could occur simultaneously 
outside of the system.
This should motivate further theoretical work to develop nonlinear 
stellar pulsation theory for more precise predictions of the 
mode behaviors in pulsating stars in general.

\begin{acknowledgements} 
Funding for the {\sl Kepler} mission is provided 
by NASA's Science Mission Directorate. We greatfully acknowledge the 
{\sl Kepler} Science Team and all those who have contributed to making 
the {\sl Kepler} mission possible. WKZ acknowledges the financial support 
from the China Scholarship Council. 
This work was supported in part by the 
Programme National de Physique Stellaire (PNPS, CNRS/INSU, France) and 
the Centre National d'Etudes Spatiales (CNES, France).
\end{acknowledgements}

\end{balance}

\newpage

\begin{table*} \caption[]{List of frequencies detected in KIC\,10139564.}
\begin{center}
\begin{tabular}{cccccccccrl}
\hline
\hline
Id.          &Frequency   &$\sigma_f$ & Period  &$\sigma_P$&Amplitude&$\sigma_A$&Phase&$\sigma_\mathrm{Ph}$&S/N &$^\dagger$Comment  \\
&             ($\mu$Hz)   &($\mu$Hz)  &(s)      &(s)       &(\%)     &(\%)                   \\
\hline
&&\\
\multicolumn{3}{l}{Multiplet frequencies:}\\
$f_{39}$ &  315.579243 & 0.000566 & 3168.776214 & 0.005687 & 0.005851 & 0.000596 & 0.2492 & 0.0516 & 9.8   & $T_{2,-1}$\\
$f_{21}$ &  315.820996 & 0.000219 & 3166.350599 & 0.002193 & 0.015155 & 0.000596 & 0.6107 & 0.0199 & 25.4  & $T_{2,0}$\\
$f_{11}$ &  316.066440 & 0.000070 & 3163.891744 & 0.000702 & 0.047276 & 0.000596 & 0.2063 & 0.0064 & 79.3  & $T_{2,+1}$\\
&&\\
$f_{27}$ &  394.027385 & 0.000342 & 2537.894669 & 0.002202 & 0.009667 & 0.000594 & 0.2589 & 0.0312 & 16.3  & $D_{1, 0}$\\
$f_{32}$ &  394.289823 & 0.000397 & 2536.205455 & 0.002555 & 0.008323 & 0.000594 & 0.5123 & 0.0363 & 14.0  & $D_{1, +1}$\\
&&\\
$f_{34}$ &  518.900359 & 0.000437 & 1927.152262 & 0.001624 & 0.007526 & 0.000592 & 0.6648 & 0.0401 & 12.7  & $T_{3, -1}$\\
$f_{28}$ &  519.151796 & 0.000352 & 1926.218898 & 0.001305 & 0.009351 & 0.000592 & 0.9059 & 0.0323 & 15.8  & $T_{3, 0}$\\
$f_{31}$ &  519.402391 & 0.000367 & 1925.289559 & 0.001360 & 0.008964 & 0.000592 & 0.5369 & 0.0337 & 15.2  & $T_{3, +1}$\\
&&\\
$f_{08}$ & 5286.149823 & 0.000053 &  189.173601 & 0.000002 & 0.064784 & 0.000614 & 0.6712 & 0.0047 & 105.4 & $Q_{1,-2}$\\
$f_{10}$ & 5286.561766 & 0.000060 &  189.158861 & 0.000002 & 0.057105 & 0.000614 & 0.4356 & 0.0053 & 92.9  & $Q_{1,-1}$\\
$f_{07}$ & 5286.976232 & 0.000038 &  189.144032 & 0.000001 & 0.088857 & 0.000614 & 0.1202 & 0.0034 & 144.6 & $Q_{1, 0}$\\
$f_{05}$ & 5287.391879 & 0.000019 &  189.129163 & 0.000001 & 0.179339 & 0.000615 & 0.3374 & 0.0017 & 291.8 & $Q_{1,+1}$\\
$f_{06}$ & 5287.805883 & 0.000029 &  189.114355 & 0.000001 & 0.119329 & 0.000615 & 0.7941 & 0.0025 & 194.2 & $Q_{1,+2}$\\
&&\\
$f_{22}$ & 5410.701146 & 0.000234 &  184.818931 & 0.000008 & 0.014871 & 0.000627 & 0.9524 & 0.0203 & 23.7  & $M_{1, 0}$\\
$f_{67}$ & 5411.143448 & 0.000958 &  184.803824 & 0.000033 & 0.003637 & 0.000627 & 0.4591 & 0.0830 & 5.8   & $M_{1, 0}$\\
$f_{13}$ & 5411.597301 & 0.000136 &  184.788325 & 0.000005 & 0.025636 & 0.000627 & 0.6770 & 0.0118 & 40.9  & $M_{1, 0}$\\
$f_{15}$ & 5412.516444 & 0.000185 &  184.756944 & 0.000006 & 0.018812 & 0.000627 & 0.8925 & 0.0160 & 30.0  & $M_{1, 0}$\\
$f_{12}$ & 5413.389096 & 0.000084 &  184.727161 & 0.000003 & 0.041339 & 0.000627 & 0.4037 & 0.0073 & 65.9  & $M_{1, 0}$\\
$f_{19}$ & 5413.814342 & 0.000222 &  184.712651 & 0.000008 & 0.015718 & 0.000627 & 0.7225 & 0.0192 & 25.1  & $M_{1, 0}$\\
&&\\
$f_{25}$ & 5570.030091 & 0.000389 &  179.532244 & 0.000013 & 0.010056 & 0.000703 & 0.5938 & 0.0300 & 14.3  & $M_{2, 0}$ \\
$f_{56}$ & 5570.484768 & 0.000964 &  179.517590 & 0.000031 & 0.004058 & 0.000703 & 0.8087 & 0.0744 & 5.8   & $M_{2, 0}$\\
$f_{61}$ & 5570.937140 & 0.001001 &  179.503013 & 0.000032 & 0.003913 & 0.000704 & 0.8254 & 0.0772 & 5.6   & $M_{2, 0}$\\
$f_{29}$ & 5571.393930 & 0.000421 &  179.488295 & 0.000014 & 0.009297 & 0.000705 & 0.5332 & 0.0325 & 13.2  & $M_{2, 0}$\\
$f_{43}$ & 5572.293674 & 0.000760 &  179.459314 & 0.000024 & 0.005168 & 0.000706 & 0.5854 & 0.0584 & 7.3   & $M_{2, 0}$\\
$f_{50}$ & 5572.728096 & 0.000902 &  179.445324 & 0.000029 & 0.004356 & 0.000707 & 0.5037 & 0.0693 & 6.2   & $M_{2, 0}$\\
$f_{45}$ & 5708.908076 & 0.000897 &  175.164845 & 0.000028 & 0.004648 & 0.000749 & 0.0571 & 0.0650 & 6.2   & $M_{2, 0}$\\
&&\\
$f_{01}$ & 5760.167840 & 0.000005 &  173.606052 & $\dots$  & 0.825132 & 0.000761 & 0.0744 & 0.0004 &1084.9 & $T_{1,-1}$\\
$f_{03}$ & 5760.586965 & 0.000008 &  173.593421 & $\dots$  & 0.554646 & 0.000761 & 0.6388 & 0.0005 & 729.3 & $T_{1,0}$\\
$f_{02}$ & 5761.008652 & 0.000007 &  173.580715 & $\dots$  & 0.567034 & 0.000761 & 0.5845 & 0.0005 & 745.5 & $T_{1,+1}$\\
&&\\
\multicolumn{3}{l}{Independent frequencies:}\\
$f_{72}$ &  892.042910 & 0.000986 & 1121.022305 & 0.001239 & 0.003319 & 0.000588 & 0.6058 & 0.0909 & 5.6   & \\
$f_{70}$ & 2212.606534 & 0.000942 &  451.955639 & 0.000192 & 0.003462 & 0.000586 & 0.0639 & 0.0872 & 5.9   & \\
$f_{37}$ & 3540.459896 & 0.000514 &  282.449182 & 0.000041 & 0.006309 & 0.000583 & 0.3837 & 0.0478 & 10.8  & \\
$f_{36}$ & 3541.431179 & 0.000477 &  282.371716 & 0.000038 & 0.006796 & 0.000583 & 0.4013 & 0.0444 & 11.7  & \\
$f_{47}$ & 4064.355754 & 0.000719 &  246.041454 & 0.000044 & 0.004526 & 0.000585 & 0.4296 & 0.0667 & 7.7   & \\
$f_{46}$ & 5048.744283 & 0.000729 &  198.069053 & 0.000029 & 0.004548 & 0.000596 & 0.9108 & 0.0664 & 7.6   & \\
$f_{41}$ & 5049.709963 & 0.000610 &  198.031176 & 0.000024 & 0.005432 & 0.000596 & 0.6548 & 0.0556 & 9.1   & \\
$f_{63}$ & 5052.604965 & 0.000879 &  197.917709 & 0.000034 & 0.003771 & 0.000596 & 0.5451 & 0.0800 & 6.3   & \\
$f_{04}$ & 5472.861431 & 0.000007 &  182.719773 & $\dots$  & 0.476915 & 0.000638 & 0.2824 & 0.0006 & 747.9 & \\
$f_{42}$ & 5709.026672 & 0.000793 &  175.161207 & 0.000024 & 0.005254 & 0.000749 & 0.0777 & 0.0575 & 7.0   & \\
$f_{53}$ & 5740.666960 & 0.001002 &  174.195787 & 0.000030 & 0.004168 & 0.000751 & 0.8817 & 0.0724 & 5.6   & \\
$f_{49}$ & 5740.807435 & 0.000946 &  174.191525 & 0.000029 & 0.004411 & 0.000751 & 0.0263 & 0.0684 & 5.9   & \\
$f_{50}$ & 5746.615392 & 0.000612 &  174.015474 & 0.000019 & 0.006881 & 0.000757 & 0.0989 & 0.0439 & 9.1   & \\
$f_{18}$ & 5747.099099 & 0.000261 &  174.000828 & 0.000008 & 0.016157 & 0.000757 & 0.8738 & 0.0187 & 21.3  & \\ 
$f_{17}$ & 5748.065581 & 0.000257 &  173.971571 & 0.000008 & 0.016414 & 0.000758 & 0.5427 & 0.0184 & 21.7  & \\
$f_{16}$ & 5749.067189 & 0.000256 &  173.941262 & 0.000008 & 0.016476 & 0.000758 & 0.1319 & 0.0183 & 21.7  & \\

\hline
\end{tabular}
\end{center}
\label{t2}
\tablefoot{
\tablefoottext{$\dagger$}{The first subscript is the identity of the multiplet and the 
second one indicates the value of $m$. The $m$-values for two 
$\ell>2$ multiplets, $M_1$ and $M_2$, are not provided as the degree $\ell$ is not known.}
}
\end{table*}

\addtocounter{table}{-1} 
\begin{table*} \caption[]{continued.}
\begin{center}
\begin{tabular}{cccccccccrl}
\hline
\hline
Id.          &Frequency   &$\sigma_f$ & Period  &$\sigma_P$&Amplitude&$\sigma_A$&Phase&$\sigma_\mathrm{Ph}$&S/N &Comment  \\
&             ($\mu$Hz)   &($\mu$Hz)  &(s)      &(s)       &(\%)     &(\%)                   \\
\hline
&&\\
$f_{44}$ & 5840.820662 & 0.000903 &  171.208818 & 0.000026 & 0.004685 & 0.000761 & 0.7872 & 0.0645 & 6.2   & \\
$f_{65}$ & 6057.645946 & 0.000970 &  165.080629 & 0.000026 & 0.003740 & 0.000652 & 0.2785 & 0.0815 & 5.7   & \\
$f_{55}$ & 6057.688799 & 0.000876 &  165.079461 & 0.000024 & 0.004142 & 0.000652 & 0.9775 & 0.0736 & 6.3   & \\
$f_{66}$ & 6106.662077 & 0.000977 &  163.755582 & 0.000026 & 0.003675 & 0.000646 & 0.6890 & 0.0822 & 5.7   & \\
$f_{60}$ & 6757.710494 & 0.000838 &  147.979112 & 0.000018 & 0.003938 & 0.000594 & 0.1752 & 0.0767 & 6.6   & \\
$f_{59}$ & 6758.215141 & 0.000835 &  147.968062 & 0.000018 & 0.003954 & 0.000594 & 0.1869 & 0.0764 & 6.7   & \\
$f_{30}$ & 7633.720521 & 0.000360 &  130.997722 & 0.000006 & 0.009090 & 0.000589 & 0.1759 & 0.0332 & 15.4  & \\
$f_{40}$ & 7634.190048 & 0.000592 &  130.989665 & 0.000010 & 0.005536 & 0.000589 & 0.3321 & 0.0546 & 9.4   & \\
$f_{64}$ & 7634.677476 & 0.000873 &  130.981302 & 0.000015 & 0.003753 & 0.000589 & 0.1106 & 0.0805 & 6.4   & \\
$f_{52}$ & 8118.752590 & 0.000768 &  123.171631 & 0.000012 & 0.004284 & 0.000591 & 0.3497 & 0.0705 & 7.2   & \\
$f_{33}$ & 8119.248304 & 0.000436 &  123.164111 & 0.000007 & 0.007538 & 0.000591 & 0.4515 & 0.0400 & 12.7  & \\
$f_{48}$ & 8496.107048 & 0.000733 &  117.700965 & 0.000010 & 0.004503 & 0.000594 & 0.8648 & 0.0744 & 7.6   & \\
$f_{71}$ & 8496.293646 & 0.000973 &  117.698380 & 0.000013 & 0.003395 & 0.000594 & 0.0896 & 0.0934 & 5.7   & \\
$f_{54}$ & 8615.236287 & 0.000795 &  116.073427 & 0.000011 & 0.004159 & 0.000594 & 0.5932 & 0.0727 & 7.0   & \\
$f_{38}$ & 8616.169582 & 0.000539 &  116.060854 & 0.000007 & 0.006128 & 0.000594 & 0.7833 & 0.0498 & 10.3  & \\
&&\\
\multicolumn{3}{l}{Linear combination frequencies:}\\
$f_{23}$ & 6076.234996 & 0.000252 &  164.575597 & 0.000007 & 0.014360 & 0.000650 & 0.7906 & 0.0210 & 22.1  &  $f_{11}+f_{01}$ \\
$f_{35}$ & 6076.408232 & 0.000510 &  164.570905 & 0.000014 & 0.007091 & 0.000650 & 0.7821 & 0.0426 & 10.9  &  $f_{21}+f_{03}$ \\
$f_{74}$ & 6076.650684 & 0.001120 &  164.564338 & 0.000030 & 0.003225 & 0.000650 & 0.5520 & 0.0937 & 5.0   &  $f_{11}+f_{03}$ \\
$f_{68}$ &  190.138219 & 0.000959 & 5259.331906 & 0.026527 & 0.003563 & 0.000614 & 0.2639 & 0.0847 & 5.8   &  $f_{01}-f_{25}$ \\
$f_{79}$ &  287.306296 & 0.001081 & 3480.605932 & 0.013093 & 0.003075 & 0.000598 & 0.9190 & 0.0982 & 5.1   &  $f_{01}-f_{04}$ \\
&&\\
\multicolumn{3}{l}{Group frequencies:}\\
$f_{20}$ & 5471.730865 & 0.000230 &  182.757527 & 0.000008 & 0.015410 & 0.000637 & 0.9397 & 0.0196 & 24.2  & G1\\
$f_{14}$ & 5944.170986 & 0.000209 &  168.232038 & 0.000006 & 0.019127 & 0.000720 & 0.1222 & 0.0158 & 26.6  & G2\\
$f_{24}$ & 6001.472409 & 0.000273 &  166.625776 & 0.000008 & 0.013532 & 0.000664 & 0.2224 & 0.0223 & 20.4  & G3\\
$f_{26}$ & 6172.852132 & 0.000353 &  161.999669 & 0.000009 & 0.009980 & 0.000633 & 0.6961 & 0.0302 & 15.8  & G4\\
$f_{09}$ & 6234.713029 & 0.000058 &  160.392306 & 0.000001 & 0.062028 & 0.000648 & 0.9590 & 0.0072 & 95.7  & G5\\    
$f_{51}$ & 6315.214679 & 0.000798 &  158.347744 & 0.000020 & 0.004312 & 0.000619 & 0.3510 & 0.0700 & 7.0   & G6\\
&&\\
\multicolumn{3}{l}{Suspected frequencies:}\\
$f_{77}$ & 4061.893709 & 0.001051 &  246.190588 & 0.000064 & 0.003097 & 0.000585 & 0.9507 & 0.0974 & 5.3   & \\
$f_{57}$ & 5838.962703 & 0.001063 &  171.263296 & 0.000031 & 0.003980 & 0.000761 & 0.6543 & 0.0758 & 5.2   & \\
$f_{62}$ & 5841.187712 & 0.001114 &  171.198059 & 0.000033 & 0.003796 & 0.000761 & 0.7222 & 0.0796 & 5.0   & \\
$f_{58}$ & 5841.581605 & 0.001068 &  171.186516 & 0.000031 & 0.003960 & 0.000761 & 0.1969 & 0.0762 & 5.2   & \\
$f_{83}$ &  345.231189 & 0.001151 & 2896.609673 & 0.009656 & 0.002878 & 0.000596 & 0.8214 & 0.1050 & 4.8   & \\
 $f_{84}$ &  345.597339 & 0.001193 & 2893.540798 & 0.009988 & 0.002777 & 0.000596 & 0.3501 & 0.1088 & 4.7   & \\
 $f_{78}$ &  345.976695 & 0.001077 & 2890.368090 & 0.008999 & 0.003076 & 0.000596 & 0.4839 & 0.0982 & 5.2   & \\
$f_{69}$ & 6106.245918 & 0.001014 &  163.766742 & 0.000027 & 0.003544 & 0.000646 & 0.6171 & 0.0852 & 5.5   & \\
$f_{75}$ & 6418.164502 & 0.001068 &  155.807786 & 0.000026 & 0.003178 & 0.000610 & 0.7474 & 0.0950 & 5.2   & \\
$f_{76}$ & 6758.687089 & 0.001049 &  147.957730 & 0.000023 & 0.003148 & 0.000594 & 0.5124 & 0.0959 & 5.3   & \\
$f_{80}$ & 6997.352981 & 0.001084 &  142.911184 & 0.000022 & 0.003003 & 0.000586 & 0.5103 & 0.1005 & 5.1   & \\
$f_{73}$ & 7633.957044 & 0.001007 &  130.993663 & 0.000017 & 0.003251 & 0.000589 & 0.2483 & 0.0930 & 5.5   & \\
$f_{81}$ & 8117.287298 & 0.001114 &  123.193866 & 0.000017 & 0.002953 & 0.000591 & 0.7915 & 0.1022 & 5.0   & \\
$f_{82}$ & 8377.175646 & 0.001119 &  119.371975 & 0.000016 & 0.002937 & 0.000591 & 0.8589 & 0.1040 & 5.0   & \\

\hline
\end{tabular}
\end{center}
\label{t2}
\tablefoot{
\tablefoottext{$\dagger$}{The first subscript is the identity of the multiplet and the 
second one denote $m$ that is discriminated by the frequency split. The $m$ for two 
$\ell>2$ multiplet $M_1$ and $M_2$ is not provided as the degree $\ell$ is not kown yet.}
}
\end{table*}

\end{document}